\documentclass[aps,superscriptaddress, showpacs,preprintnumbers, superscriptaddress, nofootinbibt,twocolumn]{revtex4}
\usepackage{eurosym}
\usepackage{dcolumn}
\usepackage{bm}
\usepackage{enumerate}
\usepackage{float}
\usepackage{epstopdf}
\usepackage{amsmath}
\usepackage{bm}
\usepackage{amsfonts}
\usepackage{amssymb}
\usepackage{graphicx}
\usepackage{alphalph,mathtools}
\usepackage{etoolbox}

\def\be{\begin{equation}}
\def\ee{\end{equation}}
\def\bea{\begin{eqnarray}}
\def\eea{\end{eqnarray}}

\begin{document}

\title{Jeans instability and turbulent gravitational collapse of
Bose-Einstein Condensate dark matter halos}

\author{Tiberiu Harko}
\email{t.harko@ucl.ac.uk}
\affiliation{Department of Physics, Babes-Bolyai University, Kogalniceanu Street,
Cluj-Napoca 400084, Romania,}
\affiliation{School of Physics, Sun Yat-Sen University, Xingang Road, Guangzhou 510275, P. R.
China,}

\begin{abstract}
 We consider the Jeans instability and the  gravitational collapse of the rotating Bose-Einstein Condensate dark matter halos, described by the zero temperature  non-relativistic Gross-Pitaevskii equation, with repulsive inter-particle interactions.  In the Madelung representation of the wave function,  the dynamical evolution of the galactic halos is described by the continuity and the hydrodynamic Euler equations, with the condensed dark matter satisfying  a polytropic equation of state with index $n=1$. By considering small perturbations of the quantum hydrodynamical equations we obtain the dispersion relation and the Jeans wave number, which includes the effects of the vortices (turbulence), of the quantum pressure and of the quantum potential, respectively. The critical scales above which condensate dark matter  collapses (the Jeans radius and mass) are discussed in detail. We also investigate the collapse/expansion of rotating condensed dark matter halos, and we find a family of exact semi-analytical solutions of the hydrodynamic evolution equations, derived by using the method of separation of variables. An approximate first order solution of the fluid flow equations is also obtained. The radial coordinate dependent mass, density and velocity profiles of the collapsing/expanding condensate dark matter halos are obtained by using numerical methods.
\end{abstract}

\date{\today}
\maketitle
\tableofcontents

\affiliation{Department of Physics, Babes-Bolyai University, Kogalniceanu Street,
Cluj-Napoca 400084, Romania}
\affiliation{School of Physics, Sun Yat-Sen University, Guangzhou 510275, People's
Republic of China}
\affiliation{Department of Mathematics, University College London, Gower Street, London
WC1E 6BT, United Kingdom}

\affiliation{Department of Physics, Babes-Bolyai University, Kogalniceanu Street,
Cluj-Napoca 400084, Romania,}
\affiliation{School of Physics, Sun Yat-Sen University, Guangzhou 510275, People's
Republic of China,}
\affiliation{Department of Mathematics, University College London, Gower Street, London,
WC1E 6BT, United Kingdom}

\section{Introduction}

After almost one hundred years of intensive study and research the properties of dark matter remain elusive. The existence of dark matter is one of the fundamental assumptions of modern cosmology and astrophysics \cite{d1,d2,d3,d4,d5,d6,d7}, and its nature is one of the most important open questions in physics. Presently, all available information on the dark sector is obtained from the study of its gravitational interactions with astrophysical systems.

 Perhaps the strongest evidence for the existence of the dark matter in the Universe comes from the study of the galactic rotation curves \cite{Rubin}.  For hydrogen clouds in stable circular orbits moving around the galactic center the rotational velocities first increase near the
center of the galaxy, thus following the standard (Newtonian) gravitational theory, but then they remain approximately constant at an asymptotic  value
of the order of $v_{tg\infty} \sim  200 - 300$ km/s. Hence the rotation curves implies the existence of a mass profile of the form $M(r) = rv^2_{tg\infty}
/G$, where $G$ is the gravitational constant. This fundamental result implies that within a distance $r$ from the center of the galaxy,
even at large distances where very little baryonic (luminous) matter can be detected, the mass  profile increases linearly with $r$. This type of  behavior can be explained by assuming the presence of a new (and exotic) mass component, interacting only gravitationally with ordinary matter, and which most likely consists of new particle(s) not (yet) included in the standard model of particle physics.  The behavior of the galactic rotation curves still provides the most convincing and compelling evidence for the existence of dark matter \cite{Persic,Read,Salucci,Haghi}.

The second important astrophysical evidence for dark matter comes  from the virial mass
discrepancy in galaxy clusters \cite{BT1}. Galaxy clusters are giant astrophysical systems formed of thousands of galaxies each, bounded together by their own gravitational interaction. The galaxies give around 1\% of the mass of the clusters, while the high temperature
intracluster gas represents around 9\% of the cluster mass. But the total masses obtained by measuring the velocity dispersions
of the galaxies exceed the total masses of all stars in the cluster by factors of the order of
200 - 400 \cite{20}. Hence, in order to explain the cluster dynamics one needs to assume the presence of dark matter, representing around
90\% of the  mass of the cluster. Another strong evidence for the presence of dark matter follows from the measurement of the temperature of the intracluster medium, since  a supplementary mass component is required to explain the determined depth of the
gravitational potential of the clusters \cite{20}. 

Several other astrophysical and
cosmological observations have also provided compelling evidence for the presence of
dark matter.  From the cosmological perspective, the recent Planck satellite measurements of the Cosmic Microwave Background
Radiation \cite{Planck1} led to the precise determination of the cosmological
parameters. These results have indicated that baryonic matter only
cannot explain the cosmological dynamics, and that the standard $%
\Lambda $ Cold Dark Matter ($\Lambda$CDM) cosmological paradigm requiring the existence of dark matter is strongly favored by observations. For a consistent interpretation of the gravitational lensing data the existence of
dark matter is also required \cite{Wegg, Munoz,
Chuda}.

Powerful observational evidences for the existence of dark matter are
yielded by the observations of a galactic cluster called the Bullet Cluster, consisting of two colliding clusters of galaxies. Due to collision of its two components that occurred in the past, in the Bullet Cluster cluster the baryonic and the dark matter components are
separated  \cite{massey2007dark}. Determinations of the cosmological parameters from the Planck
 data on the cosmic microwave background radiation did show convincingly that
the Universe consists of 74\% dark energy, 22\% non-baryonic dark matter and only 4\% baryonic matter
\cite{Planck1}.

Depending on the energy of the particles composing it, dark matter models can be classified generally into  three major types, cold, warm and hot dark
matter models, respectively,
\cite{Overduin}. For the dark matter particle(s) the main candidates  are the WIMPs (Weakly
Interacting Massive Particles) and the axions, respectively \cite{Overduin}. WIMPs are hypothetical (and yet undetected)
heavy particles, interacting through the weak force \cite{Cui,Matsumoto}.
Other popular dark matter candidates, the axions,  are bosons that were first proposed as a solution of  the strong CP (Charge+Parity) problem, which requires to explain why quantum chromodynamics  does not break the CP symmetry \cite{Mielke, Schwabe}.

There are also other approaches that attempt to interpret the observational data  without
resorting to dark matter. These explanations assume that on galactic or extra-galactic scales the
law of gravity (Newtonian or general relativistic) is modified. The earliest attempt to explain the rotation curves by modified gravity is MOND (Modified Newtonian Dynamics) theory, proposed initially in \cite{Milgrom}. Other modified theories of gravity
have also been used widely as alternative explanations to dark matter phenomenology
\cite{alt1,alt2,alt3,alt4,alt5,alt6,alt7,alt8,alt9,alt10,alt11, alt12,alt13,alt14}.  For a recent review of the dark matter problem in some modified theories of gravity see \cite{book}.

An attractive possibility for the detection of the presence of dark matter may be provided by its
possible annihilation into ordinary particles. If such physical processes really take place, then a
large number of positrons and gamma ray photons are produced, thus giving some clear observational signatures for the presence of dark matter.
This possibility may be supported by the detection of excess positron emission in our galaxy \cite{Chann1,Chann2, Chann3,Chan1,Chan2,Chan3,Chan4,Chan5}.
Hence the excess gamma-ray and positron
emissions in our galaxy could be interpreted as coming from the annihilation of dark matter with mass in the range of $m
\sim 10 - 100$ GeV \cite{Chann1,Chann2,Chann3}. For an in depth discussion of
this problem, as well as of the alternative possibilities for the interpretation
of the observational data see \cite{Chan1,Chan2,Chan3,Chan4,Chan5}.

Dark matter models can generally give  good explanations of the phenomenological (and unexpected) behaviors
of particle dynamics at the galactic and extra-galactic level,  including the constancy of the rotation curves, and the virial mass discrepancy. However, crucial conflicts do appear when one compares the results of the numerical simulations  of the theoretical models with the observations. The observational data on nearly all observed galactic rotation curves indicate that in the presence of a single pressureless dark
matter component they increase less sharply as compared to the predictions of the cosmological simulations of structure formation
in the standard $\Lambda$CDM model. Moreover, the numerical simulations display dark matter
density profiles that behave as $\rho \sim 1/r$ (a cusp) at the galactic center \cite{NFW}. On the other hand
observations of the galactic rotation curves show the existence of constant density cores
\cite{H,OH}. In dark matter physics this contradiction between theory and observations  represents the so-called core-cusp problem.
Dark matter models must also face the "too big to fail" question \cite{Boylan1,Boylan2}. The Aquarius
simulations did show that in the dark matter
halos predicted in the standard $\Lambda$CDM model the most massive subhalos are incompatible  with the observations of the dynamics
of the brightest  dwarf spheroidal galaxies of the Milky Way \cite{Boylan2}.
For the dwarf spheroidal galaxies the best-fitting hosts  have maximum velocities in the range  $12\; {\rm km/s} <  V_{max}
< 25$ km/s, while all the $\Lambda$CDM simulations give at least ten subhalos
with velocities $V_{max} > 25$ km/s. In the framework
of the $\Lambda$CDM-based models of the satellite population of the Milky
Way these observational results cannot be interpreted. The main contradiction between theory and observations is related to the predictions of the densities of the satellites, with the predicted dwarf spheroidals having dark matter halos more massive by a factor of $\sim$5 than shown by the observations.

The problems mentioned above, related to the physical properties of the dark matter, may be explained if
one extends the standard $\Lambda $CDM model by assuming that dark matter particles may have some kind of self-interaction. This model, called the Self-Interacting Dark Matter (SIDM) paradigm assumes the existence of supplementary interactions in the dark sector \cite{si1,si2,si3,si4} that may
allow momentum and energy exchange between particles that compose the dark matter halos. In these models the basic quantity describing the dark matter halo properties is the self-interaction cross section $\sigma _{DM}$ divided by the dark matter particle mass $m$, $\sigma _{DM}/m$. If the dark matter self-interactions have cross sections per mass of the same order of magnitude  as the strong nuclear force, $\sigma _{DM}/m\sim 1 {\rm g/cm ^{-2}}$, this would thermalize the inner regions of
dark matter halos where the baryonic matter is concentrated. On the other hand  for $\sigma _{DM}/m\geq 1 {\rm g/cm ^{-2}}$ on galactic scales Self-Interacting Dark Matter models can explain the uniformity as well as the diversity  of galaxy rotation curves \cite{si5,si6,si7}.
 Galaxy clusters also show the same diversity of properties of their Self-Interacting Dark Matter halos \cite{si8}.

The possibility of a complex self-interaction of dark matter particles received some observational backing from  the study of the
data obtained from the study of 72 galaxy cluster collisions. The observations did include both
`major' and `minor' mergers, and they were done by using the  Hubble and Chandra
Space Telescopes \cite{Harvey}. Important constraints on the non-gravitational forces
acting on dark matter can be obtained from the study of the collisions between galaxy.
 The analysis presented in \cite{Harvey} gave an upper
limit of  $\sigma _{DM}/m$  as $\sigma_{DM}/m < 0.47$ cm$^2$%
/g (at 95\% Confidence Level). In \cite{Jauzac} an upper limit on the self-interaction
cross-section of dark matter of $\sigma_{DM}/m<1.28$ cm$^2$/g (68\% Confidence
Level), was obtained. Different self-interacting dark matter models were investigated in \cite%
{Carlson,Laix,Saxton1,Saxton2}. In \cite{Dooley} the implications of the self-interaction of dark matter for the tidal stripping and evaporation of satellite galaxies in a Milky Way type galaxy were considered. 
The response of self-interacting dark matter halos to the
growth of galaxy potentials using numerical simulations was investigated in \cite%
{Elbert}, and a greater diversity of dark matter halo
profiles was found. A
self-interacting dark matter halo with  $%
\sigma _{DM}/m=0.1$ cm$^2$/g gives a good fit to the measured dark
matter density profile of A2667. The same halo simulated with $%
\sigma _{DM}/m=0.5$ cm$^2$/g does not produce a core profile dense enough to fit the observational data of A2667. Together with the previous findings in \cite{Harvey}, these limits point towards the result that the constraint  $\sigma
_{DM}/m\geq 0.1$ cm$^2$/g is strongly disfavored for dark matter collision velocities
greater than 1500 km/s.

Therefore, as suggested by the above observational results, we cannot reject \textit{a
priori} the possibility that dark matter is a self-interacting constituent of the Universe. Physical models of dark matter in which the fundamental particles are self-interacting may provide a better theoretical explanation of the observed phenomenology at galactic and extra-galactic scales. But from both a phenomenological perspective, and from a fundamental
theoretical and physical point of view,
\textit{the best motivated} self-interacting dark matter models
can be constructed by assuming that presently dark matter is in the form of a \textit{Bose-Einstein
Condensate}.

In the 1920s, by using  statistical physics methods,
Bose and Einstein \cite{Bose,Ein,Ein1} proposed that at temperatures smaller than a critical one, all integer spin particles will condense into
the lowest quantum state. In this phase microscopic quantum phenomena
become apparent at the macroscopic level. The Bose-Einstein
Condensation process occurs when the particles in the gas become correlated quantum mechanically,  that is,
when the de Broglie thermal wavelength turns to be
greater than the mean interparticle distance. The transition to the
condensate phase of the boson gas is initiated when the temperature $T$ is smaller
than the critical one, $T_{cr}$, which is given by the expression \cite{Dalfovo, Pita, Pethick, ZNG}
\begin{equation}\label{Ttr}
T_{cr}=\frac{2\pi\hbar^2\rho_{cr}^{2/3}}{ \zeta^{2/3}(3/2)m^{5/3}k_B},
\end{equation}
where $\rho_{cr}$ is the
critical transition density, $m$ is the mass of the particle in the boson gas,  $k_B$ is Boltzmann's constant, and $\zeta $
denotes the Riemann zeta function, respectively.

The experimental creation of
Bose-Einstein Condensates was first realized in dilute alkali gases
in 1995 by cooling a dilute vapor of approximately two thousand rubidium-87 atoms to below 170 nK, using a combination of laser cooling and magnetic evaporative cooling \cite{exp1,exp2,exp3}. The presence of a Bose-Einstein Condensate in a bosonic system is indicated, from a physical and experimental  point of view, by the appearance in both coordinate and momentum space distributions of the particles of sharp peaks.

Presently, the only evidence for the existence of Bose Einstein Condensates on a microphysical scale appeared in laboratory
experiments, which involve a very small scale. On the other hand, the possibility of
the existence of some forms of bosonic condensates in the cosmic environment cannot
be excluded a priori. Due to their
superfluid properties, in high density general relativistic objects, like, for example, neutron
or quark stars, the neutrons or the quarks could form Cooper pairs, which, once the temperature or density reach their critical values,  would eventually condense. Bose-Einstein Condensate stars may have maximum central densities of the order of $%
0.1-0.3\times 10^{16}$ g/cm$^3$, minimum radii in the range of 10-20 km, and maximum masses of the
order of  $2M_{\odot}$, respectively. The study of their interesting physical and astrophysical properties is presently an active field of research \cite{starsm1, stars0, stars1,stars2,stars3,stars4,stars5,stars6,stars7, stars8, stars9}.

Since dark matter is assumed to be a gas of bosonic particles, it naturally follows that it may have experienced a phase transition during its cosmological history, and presently it is in the form of a Bose-Einstein Condensate. The possibility that dark matter may be in the form of a Bose-Einstein Condensate was initially suggested in \cite{early1}, and then rediscovered/reinvestigated,
in \cite{early2, early3,early4,early5, early6,early7, early8, early9a,
silverman2002dark, rotha2002vortices, early9}. The systematic investigation of the
physical and astrophysical properties of the Bose-Einstein Condensate dark matter halos, based on the non-relativistic
Gross-Pitaevskii equation \cite{GP1,GP2} in the presence of an external confining gravitational
potential, was started in \cite{BoHa07a}. A significant simplification of the
mathematical and physical formalism required for the description of the gravitationally bounded Bose-Einstein Condensates can be obtained
by introducing the Madelung representation of the wave function. The Madelung representation
provides an equivalent description of the Gross-Pitaevskii equation in a form similar to the hydrodynamic fluid flow equations in classical mechanics, that is,  a continuity equation, and an Euler type equation, respectively. With the help of the Madelung hydrodynamical representation one obtains  the fundamental result that Bose-Einstein Condensed dark matter can be described as a non-relativistic, Newtonian  gas, in a gravitational trapping potential, with the pressure and density obeying a polytropic equation of state, with polytropic index $n=1$ \cite{Horedt}. From this result it follows that the radius $R$ of the zero temperature Bose-Einstein Condensate dark matter
halo is given by $R=\pi \sqrt{\hbar ^{2}a/Gm^{3}}$, where $a$ is the scattering length \cite{BoHa07a}. The total
mass $M$  of the condensate dark matter halo is obtained as $M=4\pi
^2\left(\hbar ^2a/Gm^3\right)^{3/2}\rho _c=4R^3\rho _{c}/\pi $, where $%
\rho _{c}$ is the central density of the galactic halo.  The mass of the dark matter particle  satisfies
 a particle mass-galactic radius relation given by \cite{BoHa07a}
\begin{eqnarray}  \label{mass}
m &=&\left( \frac{\pi ^{2}\hbar ^{2}a}{GR^{2}}\right) ^{1/3}\approx
6.73\times 10^{-2}\times \left[ a\left( \mathrm{fm}\right) \right] ^{1/3}\times \nonumber\\
&&\left[ R\;\mathrm{(kpc)}\right] ^{-2/3}\;\mathrm{eV}.
\end{eqnarray}

The effectiveness of the Bose-Einstein Condensate  dark matter model was checked through the fitting of the Newtonian and general relativistic tangential velocity equation to a sample of observed rotation curves data of dwarf and low surface brightness galaxies, respectively.

The condensate dark matter halo may not necessarily be at zero temperature. In
\cite{HaM} the thermal correction to the dark matter density profile where obtained.  An important result on the Bose-Einstein Condensate dark matter halos is that their density profiles  generally indicate the presence of an
enlarged core, whose presence is due to the strong interaction
between dark matter particles \cite{Har1}. The investigations of the properties of the Bose - Einstein Condensate
dark matter on  cosmological and astrophysical scales is a very
interesting and active field of research \cite{inv0, inv1,
inv2,inv3,inv4,inv5,inv6,inv7,inv8,inv9,inv10,inv11,inv12,inv13,inv14,inv15,inv16,inv17,inv18,inv19,inv20,inv21,inv22,inv23,inv24,inv25,inv26, inv27,inv28,inv29,inv30,inv31,inv32,inv33,inv34,inv35, inv36,inv37, inv38,inv39,inv40,inv41, inv41a, inv42, inv43, inv44,inv45}. Properties of the Fuzzy Dark Matter, a particular theoretical form  of dark matter, assumed to be formed of an extremely light boson ($%
m\sim 10^{-22}$ eV), with a de Broglie wavelength of the order of $\lambda \sim 1$
kpc, were investigated in \cite{Hui}.

While the static properties of the Bose-Einstein Condensate dark matter halos have been extensively investigated, their rotational characteristics have received less attention. The presence of vortices in a self gravitating Bose-Einstein Condensate dark matter halo,
consisting of ultra-low mass scalar bosons, was investigated in
\cite{early9a}. Rotation of the dark matter may induce a harmonic trap potential for
vortices. In \cite{inv3} a detailed study of the vortices in rotating Bose-Einstein Condensate dark matter halos
was performed, and strong bounds for the shape and quantity of vortices in the halo, for interaction strength, for the critical rotational velocity for the nucleation of vortices, and for the boson mass were found.
 In \cite{inv4},  by assuming that a vortex lattice forms, the effects of rotation on a superfluid Bose-Einstein Condensate dark matter halo
were investigated. On the rotation curves sub-structures
similar to some observations in spiral galaxies may form. The equilibrium properties of self-gravitating, rotating  Bose-Einstein Condensate haloes, which satisfy
the Gross-Pitaevskii-Poisson equations were studied in \cite{inv15}.
For a wide range of the Bose-Einstein Condensate dark matter physical parameters vortices are generated. On the other hand,
vortices cannot appear for a vanishing self-interaction, and they form when the self-interaction between dark matter particles is strong enough.

One of the fundamental concepts in modern astrophysics and cosmology is that of gravitational
instability, initially discussed by Jeans \cite{Jeans}. Its importance is related to the prospect of
estimating the scale of the condensations that may occur in an extended gaseous
medium under the influence of small perturbations.  The possibility of such an estimation will provide at least qualitative information
regarding the formation of stars and of galaxies from an original cosmic medium. The main result of the original analysis by Jeans was that a
self-gravitating infinite uniform gas at rest should be unstable against
small perturbations proportional to $\exp \left[ i\left( \vec{k}\cdot \vec{r}%
-\omega t\right) \right] $. The lineariazation of the equations of the ideal
hydrodynamics and the Poisson equation results in the well-known dispersion
relation $\omega ^{2}=c_{s}^{2}k^{2}-4\pi G\rho $, where  $%
c_{s}=\left( \gamma k_{B}T/m\right) ^{1/2}$ is the adiabatic sound velocity, $\rho $ is the density, $T$ is the gas temperature, and $%
\gamma =5/3$ is the ratio of specifics heats, respectively. When $\omega
^{2}$ becomes negative, an instability arises once the perturbation
wavelength $\lambda =2\pi /k$ exceeds the critical value $\lambda _{J}=c_{s}%
\sqrt{\pi /G\rho }$, $\lambda >\lambda _{J}$. Thus, an originally uniform
gas, due to the instability, should break into massive components with
characteristic size of the order of $\lambda _{J}$ \cite{BT1}. The effects of the rotation and of other physical effects on the stability of a self-gravitating medium were considered in \cite{Chand1, Chand2, Chand2a, Chand3,Chand4,Chand5,Chand6,Chand7, Chand8, Chand9, Chand10}.
The kinetic theory of the Jeans instability was developed in \cite{kin1}-\cite{kin5}, mainly using
methods from plasma physics. For a detailed discussion of the Jeans instability and its role in astrophysics see \cite{BT1}.

An interesting property of the Bose-Einstein Condensation processes is represented by the collapse and the ensuing explosion of the
condensates \cite{CollBEC1}.  Near a Feshbach resonance
the atomic scattering length $a$ can be changed, by adjusting an external magnetic field, over a large
range. Once the sign of the scattering length is changed,
a repulsive condensate of $^{85}$ Rb atoms is transformed into
an attractive one that subsequently reaches  a collapsing and an exploding phase. The collapse of a Bose-Einstein Condensate was investigated  by using the semi-classical Fokker-Planck equation for a gas of free bosons in \cite{CollBEC2, CollBEC3, CollBEC4}, and, for a $1/r^b$ type potential, in \cite{Lush}.

The mechanisms of the gravitational collapse of the Bose-Einstein Condensate dark matter halos was studied in \cite{inv23}, by using a variational
approach, and by choosing an appropriate trial wave function. This approach allows the reformulation of the Gross-Pitaevskii
equation with spherical symmetry as Newton’s equation of motion for a particle in an effective
potential, which is determined by the zero point kinetic energy, the gravitational energy, and the
particles interaction energy, respectively. The velocity of the condensate is proportional to the
radial distance, with a time dependent proportionality function. The
collapse of the condensate ends with the formation of a stable configuration, corresponding to the
minimum of the effective potential. The obtained results did show that the gravitational
collapse of the condensed dark matter halos can lead to the formation of stable astrophysical systems
with both galactic and stellar sizes.

It is the goal of the present paper to investigate the Jeans stability and the collapsing properties of the Bose-Einstein Condensate dark
matter halos in the presence of vortices, induced by the rotation of the halo. Rotation is a general
feature of galactic dynamics, probably generated by some physical instability
processes in the early Universe. To describe the Bose Einstein
Condensate galactic dark matter halos as a multi-particle bosonic system we adopt an effective approach based on the Gross-Pitaevskii equation \cite{BoHa07a}. The Gross-Pitaevskii equation gives
an effective mean-field description of the gravitationally confined dark matter halo.
The mathematical analysis of the condensed dark matter halos is greatly simplified
by the introduction of the Madelung representation of the wave function, which allows the
description of the dark matter in terms of the equations of the classical fluid dynamics, the continuity and the Euler equation, respectively. In this approach dark matter halos can de described as  fluid structures obeying a polytropic equation of
state, with polytropic index $n=1$. The Euler evolution equation also contains the quantum force, derived from the quantum potential, and which represents a purely quantum effect.  The assumptions of the rotation for the condensate dark matter halo leads to the necessity of taking into account the presence of quantized vortices. A vortex is an excitation of the bosonic system,  and hence it is a state whose energy is higher
than that of the ground-state.

In order to investigate the Jeans stability of Bose-Einstein Condensate dark matter halos we consider the perturbation of the hydrodynamic flow equations with respect to an appropriately chosen ground state, and we perform a local linear stability analysis, which includes the effects of the quantum pressure, of the quantum potential and of the rotation, as well as the self gravity of the system. The dispersion relation for the propagation of the instabilities in the system is obtained, and the Jeans wave number for the rotating Bose-Einstein Condensate dark matter halo is obtained. Once the Jeans wave number is known, we can immediately obtain the Jeans radius and mass,  which give the critical length and mass scales above which condensate dark matter  collapses. In the approximation when rotation and quantum force are ignored,  the Jeans radius and mass coincide with the radius and mass of the static condensate. The effects of the rotation are also considered, and the Jeans radius and mass are also obtained for rotation dominated dark matter halos. Similar results can be obtained in the framework of the Thomas-Fermi approximation, by requiring that the total energy of the system is minimal.

Once the radius and mass of the Bose-Einstein Condensate dark matter halo exceeds the Jeans limits for the critical stability, gravitational collapse or expansion follows. In order to study the collapse or the expansion of the condensate we derive first a semi-analytical solution to the spherical
hydrodynamic equations governing the time and space evolution of a Bose-Einstein Condensate dark matter halo. First we reduce the system of the coupled nonlinear partial equations describing the evolution of the galactic halo to a single, strongly nonlinear partial differential equation for the mass $M=M(r,t)$ of the condensate. Then we factorize the mass function into products of a time-dependent factor and another factor
depending only on the spatial variable $r$. This factorization means that both spatial and temporal
profiles of the gravitational, mechanical and thermodynamic variables of the halo have a universal behavior, once they are factorized
appropriately. To study the spatial profiles of the collapsing/expanding rotating dark matter halos we use approximate and numerical methods. A first order approximative solution of the mass function is explicitly obtained. The general mass, density and velocity spatial profiles are obtained by numerically integrating the mass equation. In relation with the general properties of the gravitational collapse we also show that homologous solutions to the hydrodynamic  equations describing the time evolution of a Bose-Einstein
condensate dark matter, wherein thermodynamic variables
factorize into products of a time-dependent factor and another factor
depending only on the scaled spatial variable $\xi \equiv r/R(t)$, {\it do not exist}. This factorization means that spatial
profiles of the thermodynamic variables remain time invariant. Since Bose-Einstein Condensate dark matter halos do not have this property, it follows that the dynamics of the collapse of a condensed galactic halo
is very different from other types of gravitational collapse, like, for example, the pre-supernova stellar core collapse.

The present paper is organized as follows. The basic properties of the Bose-Einstein Condensate dark matter halos are presented in Section~\ref{sect1}, where, with the use of the Madelung representation,  the evolution equations of the condensed galactic halos are formulated in terms of the continuity and Euler equations of classical hydrodynamics. The Jeans instability of condensed dark matter halos confined by their own gravitational field is discussed in Section~\ref{sect2}, by fully taking into account the effects of the rotation and the quantum effects. The limiting cases of the general dispersion relation are considered in detail. The time evolution of the Bose-Einstein Condensate dark matter halos is investigated in Section~\ref{sect3}, and it is shown that the hydrodynamic equations describing the dynamics of a rotating polytropic gas with polytropic index $n=1$ do admit a separable solution, in which all physical quantities can be expressed as products of two functions, one depending on time only, and the second a function of the radial coordinate only. An approximate first order solution of the evolution equations is also obtained. The spatial dependence of the physical quantities is investigated by using numerical methods. We discuss and conclude our results in Section~\ref{sect4}. In Appendix~\ref{appa} we explicitly show that the hydrodynamic evolution equations describing a polytropic fluid with polytropic index $n=1$ do not admit self-similar homologous solutions.

\section{Gravitationally confined Bose-Einstein dark matter halos}\label{sect1}

Generally, Bose-Einstein condensation processes take place in a Bose gas with particle number density $n$
when the thermal de Broglie wave length $\lambda _{dB}=\sqrt{2\pi \hbar ^2/mk_BT}$,
 exceeds the mean inter-particle distance $n^{1/3}$. Then, as a result,  the wave packets percolate in
space. The critical condensation temperature can then be obtained qualitatively as  $T\leq2\pi\hbar ^2n^{2/3}/mk_B$
\cite{BoHa07a}. Under the assumption of an adiabatic cosmological expansion of the
Universe, the temperature dependence of the number density of the particle
is $T\propto n^{2/3}$. Hence Bose-Einstein condensation occurs if the mass
of the particle satisfies the condition $m<1.87$ eV \cite{Bos}, and therefore particles satisfying this mass limit
 could Bose-Einstein condense, and form large scale cosmological or astrophysical structures.

Due to the low temperature of the Bose-Einstein Condensates, their physical
properties can be understood within the so called mean-field approximation.
The success of the mean field approach is determined by the dilute nature of
the Bose gases. We can always write the Bose field operator as a sum of the
condensate wave function and an operator describing the non-condensed
bosons, so that $\hat{\Psi }\left(\vec{r},t\right)=\Psi \left(\vec{r}%
,t\right)+\hat{\Psi}^{\prime }\left(\vec{r},t\right)$, where $\Psi \left(%
\vec{r},t\right)\equiv \left<\hat{\Psi }\left(\vec{r},t\right)\right>$ is
the average value of $\hat{\Psi }\left(\vec{r},t\right)$, and $\hat{\Psi}%
^{\prime }\left(\vec{r},t\right)$ represents the fluctuations in the system.
The single-particle density matrix $\rho $ is given by $\rho \left(\vec{r},%
\vec{r}\;^{\prime }\right)=\left<\hat{\Psi }^{+}\left(\vec{r}\right)\hat{%
\Psi }\left(\vec{r}\;^{\prime }\right)\right>$, where $\hat{\Psi }^{+}\left(%
\vec{r}\right)$ is the field operator creating a particle at a point $\vec{r}
$, and $\hat{\Psi }\left(\vec{r}\;^{\prime }\right)$ is the field operator
annihilating a particle at $\vec{r}\;^{\prime }$. In a dilute Bose gas close
to $T = 0$, one can neglect with a very good approximation the non-condensed
bosons $\hat{\Psi}^{\prime }\left(\vec{r},t\right)$. In this case the
mean-field order parameter is given exactly by the quantum mechanical wave
function $\Psi \left(\vec{r},t\right)$, with well defined phase. Hence, the
zero temperature dynamics of a Bose gas consisting of confined weakly
interacting particles is described by a mean-field macroscopic wave function
$\Psi$. The wave function of the condensate behaves like a complex order
parameter whose absolute value and phase contains all relevant information
about the Bose-Einstein Condensate system, and satisfies a nonlinear Schr%
\"{o}dinger equation, called the Gross-Pitaevskii equation \cite{GP1, GP2}.

Hence, gravitationally confined Bose-Einstein Condensate dark matter halos
are described by the coupled Gross-Pitaevskii and Poisson equations for the
condensate wave function $\Psi $, with a quartic non-linear term, given by
\cite{Dalfovo, Pita, Pethick, ZNG}
\begin{eqnarray}  \label{sch}
i\hbar \frac{\partial }{\partial t}\Psi \left( \vec{r},t\right) =\hat{H}\Psi
(\vec{r},t)&=&\Bigg[ -\frac{\hbar ^{2}}{2m}\nabla ^{2}+mV_{ext}(\vec{r},t)+
\notag \\
&&U_{0}|\Psi (\vec{r},t)|^{2}\Bigg] \Psi (\vec{r},t),
\end{eqnarray}
where $U_{0}=4\pi \hbar ^{2}a/m$, $a$ is the coherent scattering length
(defined as the zero-energy limit of the scattering amplitude $f_{scat}$), $%
m $ is the mass of the condensate particle, and $V_{ext}$ is the external
potential. In the following we assume that the exterior potential $V_{ext}(%
\vec{r},t)$ is the gravitational potential $V_{grav}$, $V_{ext}(\vec{r}%
,t)=V_{grav}\left( \vec{r},t\right) \equiv \phi \left( \vec{r},t\right) $.
For a single component condensate dark matter halo $\phi \left( \vec{r}%
,t\right) $ satisfies the Poisson equation
\begin{equation}
\nabla ^{2}\phi (\vec{r},t)=4\pi G m\left\vert \Psi \left( \vec{r},t\right)
\right\vert ^{2}=4\pi G\rho (\vec{r},t),  \label{poi}
\end{equation}%
where $\rho =mn\left( \vec{r},t\right) =m\left\vert \Psi \left( \vec{r}%
,t\right) \right\vert ^{2}$ is the mass density inside the Bose-Einstein
condensate, $n\left( \vec{r},t\right) =\left\vert \Psi \left( \vec{r}%
,t\right) \right\vert ^{2}$ is the particle number density, and $G$ is the
gravitational constant, respectively. The probability density $\left\vert
\Psi \left( \vec{r},t\right) \right\vert ^{2}$ is normalized according to $%
\int_{V}{n\left( \vec{r},t\right) d^{3}{\vec{r}}}=\int_{V}{\left\vert \Psi
\left( \vec{r},t\right) \right\vert ^{2}d^{3}\vec{r}}=N$, where $N$ is the
total particle number in the dark matter halo, which can be obtained by
integrating the norm of the wave function over the entire volume $V$ of the
Bose-Einstein Condensate.

The time dependent Gross-Pitaevski Eq. (\ref{sch}) can be derived from the
action principle $\delta \int_{t_{1}}^{t_{2}}Ldt=0$, where the Lagrangian $L$
is given by \cite{var}
\begin{equation}
L=\int \frac{i\hbar }{2}\left( \Psi ^{\ast }\frac{\partial \Psi }{\partial t}%
-\Psi \frac{\partial \Psi ^{\ast }}{\partial t}\right) dV-E,  \label{lagr}
\end{equation}%
where $E=\int \varepsilon dV$ is the total energy, with the energy density $%
\varepsilon $ given by
\begin{equation}
\varepsilon =\frac{\hbar ^{2}}{2m}\left\vert \nabla \Psi \right\vert
^{2}+m\phi \left\vert \Psi \right\vert ^{2}+\frac{u_{0}}{2}\left\vert \Psi
\right\vert ^{4}.  \label{energy}
\end{equation}

If one multiplies Eq.~(\ref{sch}) by $\Psi ^{\ast }$ and subtracts the
complex conjugate of the resulting equation, one arrives at the equation
\begin{equation}
\frac{\partial \left| \Psi \right| ^{2}}{\partial t}+\nabla \cdot \left[
\frac{\hbar }{2im}\left( \Psi ^{\ast }\nabla \Psi -\Psi \nabla \Psi ^{\ast
}\right) \right] =0.  \label{cont0}
\end{equation}

Eq.~(\ref{cont0}) has the form of a continuity equation for the particle
density, and can be written as a continuity equation,
\begin{equation}  \label{cont}
\frac{\partial \rho }{\partial t}+\nabla \cdot \vec{j} =0,
\end{equation}%
where the momentum $\vec{j}$ of the condensate is defined by
\begin{equation}  \label{mom}
\vec{j}=\rho \vec{v}=\frac{\hbar }{2i}\left( \Psi ^{\ast} \nabla \Psi -\Psi
\nabla \Psi ^{\ast} \right) ,
\end{equation}%
where $\vec{v}$ is the condensate velocity. By representing the wave
function as \cite{Dalfovo, Pita, Pethick, ZNG}
\begin{equation}  \label{8}
\Psi \left( \vec{r},t\right) =\varphi \left( \vec{r},t\right) \exp \left[ iS
\left( \vec{r},t\right) \right],
\end{equation}
where $S \left( \vec{r},t\right)$ is the \textit{phase} of the wave
function, we obtain
\begin{equation}
\vec{j}=m\varphi ^2\left(\vec{r},t\right)\frac{\hbar}{m}\nabla S\left(\vec{r}%
,t\right),
\end{equation}
which gives
\begin{equation}
\varphi \left(\vec{r},t\right)=\sqrt{\frac{\rho\left(\vec{r},t\right)}{m}},
\end{equation}
and
\begin{equation}  \label{11}
\vec{v}=\frac{\hbar}{m}\nabla S\left(\vec{r},t\right),
\end{equation}
respectively.

By substituting the wave function (\ref{8}) into the Gross-Pitaevskii
equation (\ref{sch}), it follows that it can be decomposed into two
independent equations, the continuity equation (\ref{cont}), and the
Hamilton-Jacobi equation
\begin{equation}
-\hbar \frac{\partial S}{\partial t}=-\frac{\hbar ^2}{2m\varphi}\nabla
^2\varphi +\frac{1}{2}m\vec{v}^2+m\phi +U_0\varphi ^2.
\end{equation}

By taking the gradient of the above equation we obtain
\begin{eqnarray}  \label{13}
\frac{d\vec{v}}{dt}&=&\frac{\partial \vec{v}}{\partial t}+\left(\vec{v}\cdot
\nabla\right)\vec{v}=-\frac{1}{\rho} \nabla p-m\nabla \phi +  \notag \\
&&\frac{1}{m}\nabla \left(\frac{\hbar ^2}{2m}\frac{\nabla ^2\sqrt{\rho}}{%
\sqrt{\rho}}\right).
\end{eqnarray}
where $p=u_0\rho ^2$ is the quantum pressure, with $u_0=2\pi \hbar ^2a/m^3$,
while the last term in Eq.~(\ref{13}) is the quantum force, derived from the
quantum potential $V_Q=-\left(\hbar ^2/2m\right)\nabla ^2 \sqrt{\rho}/\sqrt{%
\rho}$. By taking into account the mathematical identity
\begin{equation}
\left(\vec{v}\cdot \nabla\right)\vec{v}=\frac{1}{2}\nabla \vec{v}^{\;2}-\vec{%
v}\times \left(\nabla \times \vec{v}\right),
\end{equation}
we can rewrite the Euler Eq.~(\ref{13}) as
\begin{eqnarray}  \label{15}
&&\frac{\partial \vec{v}}{\partial t}+\frac{1}{2}\nabla \vec{v}^{\;2}=-\frac{%
1}{\rho} \nabla p-m\nabla \phi +  \notag \\
&&\frac{1}{m}\nabla \left(\frac{\hbar ^2}{2m}\frac{\nabla ^2\sqrt{\rho}}{%
\sqrt{\rho}}\right)+\vec{v}\times \left(\nabla \times \vec{v}\right).
\end{eqnarray}

The integration of Eq.~(\ref{cont}) over the volume $V$ of the condensate,
and by using use of the Gauss theorem, leads immediately to the equation of
the particle number conservation, which can be formulated as
\begin{equation}  \label{9}
\frac{\partial }{\partial t}N+\frac{1}{m}\int_{S}{\rho \vec{v}\cdot \vec{n}dS%
}=0,
\end{equation}%
From Eq.~(\ref{9}) it follows that the time variation of the particle number
can be due only to the loss or gain of the dark matter particles moving in
or out of the condensate through the surface $S$ that encompasses the total
volume $V$. For a static condensate, $\vec{v}\equiv 0$, and $\rho (R)\equiv
0 $, where $R$ is the radius of the condensate, a condition which implies
that the particle flux $\vec{j}(R)=\rho (R)\vec{v}\equiv 0$. Hence, from
Eq.~(\ref{9}), it follows that the total particle number in the
Bose-Einstein Condensate dark matter halo at zero temperature is a constant,
$N=\mathrm{constant}$.

For superfluid for which the phase of the wave function is \textit{%
nonsingular}, the continuity and Euler Eqs.~(\ref{cont}) and (\ref{15})
represent a \textit{potential (irrotational) flow}, since the condition $%
\nabla \times \vec{v}=0$ is always satisfied, due to the definition (\ref{11}%
) of the fluid velocity.

The vorticity can be present in a Bose-Einstein Condensate due to \textit{%
the phase singularity lines} \cite{vor0, vor1,vor2}. Since the phase of the wave function is
defined within a factor of $2\pi$ only, Eq.~(\ref{11}, defining the
condensate velocity, implies that the circulation of the velocity field
along a closed contour must be quantized in units of $\hbar/m$, according to
the rule \cite{vor0, vor1,vor2}
\begin{equation}
\Gamma =\oint {\vec{v}\cdot d\vec{l}}=n\frac{\hbar}{m},
\end{equation}
where $n$, called the topological charge of the flow, is an integer number.
It also represents the winding number of the phase of the wave function
along the contour. To obtain a non-zero circulation, the contour must winds
around a line of zero density, along which the phase, as well as the
velocity field, are no longer defined. The nucleation of the quantized
vortices is a powerful experimental evidence for the existence of the
macroscopic wave function describing the Bose-Einstein Condensate. In
laboratory experiments the observation of the quantized vortices is quite
difficult due to the smallness of the size of the vortex core. This size is
of the order of the so called \textit{healing length} \cite{Pethick}
\begin{equation}  \label{xi}
\xi _h= \frac{\hbar}{mv_s},
\end{equation}
where $v_s$ is the velocity of the sound in the condensate (for $\mathrm{He}%
^4$, the healing length is of the order of a few angstroms).

In the case of the presence in the condensate of a sufficiently large number
of vortices we can write \cite{vor0, vor1,vor2}
\begin{equation}  \label{19a}
\nabla \times \vec{v} = 2\vec{\Omega},
\end{equation}
where $\vec{\Omega}$ is \textit{the macroscopic angular velocity of the
condensate}, given by
\begin{equation}  \label{20}
\left| \vec{\Omega}\right| =\pi \frac{\hbar }{m}n_V,
\end{equation}
where $n_V$ is \textit{the areal vortex number density}, defined as $n_V =
N_V /A_{\perp}$, and obtained by assuming that the singular velocity fields
of the vortices are distributed uniformly in the plane of rotation with area
$A_{\perp}$. As we move away from the vortex, the velocity slowly decreases.
If we move towards the vortices then the superfluid density tends to zero.

This approximation is called the approximation of the \textit{distributed
vorticity}, and it has been used very successfully to describe the dynamics
of Bose-Einstein Condensates containing vortex lattices. In the distributed
vorticity approximation we introduce a rotational component in the velocity
field of the condensate, so that we can write \cite{vor0, vor1,vor2}
\begin{equation}
\vec{v}=\vec{v}_I+\vec{v}_R,
\end{equation}
where $\vec{v}_I$ represents the irrotational component, while $\vec{v}_R =%
\vec{\Omega }\times \vec{r}$.

\section{The Jeans instability for turbulent Bose-Einstein Condensate dark
matter halos}\label{sect2}

To study the astrophysical and cosmological implications of the presence of
vortex turbulence in the condensate Bose-Einstein dark matter, and its
impact on structure formation, we assume that as a result of an
external/internal perturbation the Bose-Einstein Condensate dark matter
structure becomes unstable, and evolves in time like a non-relativistic,
dissipationless fluid. The dynamical evolution of the condensed dark matter
halo is described by the continuity equation, the hydrodynamical Euler
equation and the Poisson equation, respectively, which can be written as
\begin{equation}
\frac{\partial \rho }{\partial t}+\nabla \cdot \left( \rho \vec{v}\right) =0,
\end{equation}%
\begin{equation}
\frac{\partial \vec{v}}{\partial t}+\frac{1}{2}\nabla \vec{v}^{2}=-\frac{1}{%
\rho }\nabla p+\vec{g}+\frac{1}{m}\nabla \left( \frac{\hbar ^{2}}{2m}\frac{%
\nabla ^{2}\sqrt{\rho }}{\sqrt{\rho }}\right) +2\vec{v}\times \vec{\Omega},
\label{hydr1}
\end{equation}%
\begin{equation}
\nabla \times \vec{g}=0,\nabla \cdot \vec{g}=-4\pi G\rho ,  \label{hydr2}
\end{equation}%
where $\vec{g}=-\nabla \phi $ is the gravitational acceleration. We take as
the initial (unperturbed) state of the condensed dark matter the state
characterized by the absence of the "real" gravitational interactions, $\vec{%
g}=\vec{g}_{0}=0$, of the hydrodynamical flow, $\vec{v}=\vec{v}_{0}=0$, and
by constant values of the density and pressure, $\rho =\rho _{0}$, $p=p_{0}$%
, respectively. Moreover, we assume an initial rotation of the dark matter
cloud, so that $\vec{\Omega}(0)=\vec{\Omega}_{0}\neq 0$. The instability in
the dark matter halo results in the appearance of the gravitational
interaction in the system, and of the small perturbations of the
hydrodynamical quantities, which in the first order linear approximation can
be represented as
\begin{eqnarray}
\rho &=&\rho _{0}+\delta \rho ,\;p=p_{0}+\delta p,\;\vec{\Omega}=\vec{\Omega}%
_{0}+\delta \vec{\Omega},  \notag \\
\vec{v} &=&\vec{v}_{0}+\delta \vec{v},\;\vec{g}=\vec{g}_{0}+\delta \vec{g},
\end{eqnarray}%
where we assume $-1<<\delta \rho /\rho _{0}<<1$, $-1<<\delta p/p_{0}<<1$, $%
\left|\delta \vec{g}\right|/\left|\vec{g}_{0}\right|<<1$, and $\left|\delta
\vec{\Omega}\right|/\left|\vec{\Omega }_{0}\right|<<1$, respectively.
Therefore in the first order linear approximation Eqs.~(\ref{hydr1}) and (%
\ref{hydr2}) take the form
\begin{equation}
\frac{\partial \delta \rho }{\partial t}+\rho _{0}\nabla \cdot \delta \vec{v}%
=0,  \label{19}
\end{equation}%
\begin{equation}  \label{hydr3}
\frac{\partial \delta \vec{v}}{\partial t}=-\frac{v_{s}^{2}}{\rho _{0}}%
\nabla \delta \rho +\delta \vec{g}+\frac{\hbar ^{2}}{4m^{2}\rho _{0}}\Delta
\nabla \delta \rho +2\delta \vec{v}\times \vec{\Omega}_{0},
\end{equation}%
\begin{equation}
\nabla \times \delta \vec{v}=-2\vec{\Omega}_{0}\frac{\delta \rho }{\rho _{0}}%
,\;\nabla \times \delta \vec{g}=0,\;\nabla \cdot \delta \vec{g}=-4\pi
G\delta \rho ,  \label{28}
\end{equation}%
where we have introduced the adiabatic speed of sound $v_{s}$ in the
unperturbed dark matter fluid, defined as
\begin{eqnarray}
v_{s} &=&\left. \sqrt{\frac{\delta p}{\delta \rho }}\right\vert _{\rho =\rho
_{0}}=\left. \sqrt{\frac{\partial p}{\partial \rho }}\right\vert _{\rho
=\rho _{0}}=\sqrt{2u_{0}\rho _{0}}=\sqrt{\frac{4\pi \hbar ^{2}a}{m^{3}}\rho
_{0}}=  \notag \\
&&1.57\times 10^{7}\times \left( \frac{a}{10^{-3}\;\mathrm{fm}}\right)
^{1/2}\left( \frac{\rho _{0}}{10^{-24}\;\mathrm{g/cm^{3}}}\right)
^{1/2}\times  \notag \\
&&\left( \frac{m}{\mathrm{meV}}\right) ^{-3/2}\;\mathrm{cm/s}.
\end{eqnarray}

To obtain the first of the equations in (\ref{28}) we have taken the
variation of Eq.~(\ref{19}) as follows: $\nabla \times \left( v_{0}+\delta
\vec{v}\right) =\nabla \times \delta \vec{v}=2\delta \vec{\Omega}$, $\delta
\left\vert \vec{\Omega}\right\vert =\left( \pi \hbar /m\right) n_{V}\left(
\delta n_{V}/n_{V}\right) =-\left\vert \vec{\Omega}\right\vert _{0}\delta
\rho /\rho _{0}$, where we have assumed that $\delta n_{V}/n_{V}=\delta \rho
_{V}/\rho _{V}=-\delta \rho /\rho _{0}$, that is, the relative variation of
the condensate areal density $\rho _{V}=mn_{V}$ is inversely proportional to
the variation of the density of the condensate normalized with respect to
the initial density (the higher the mass density of the vortices, the lower
the mass density of ordinary matter in the condensate).

After taking the partial derivative with respect to the time of the
continuity equation (\ref{19}), and applying the $\nabla $ operator to Eq.~(%
\ref{hydr3}), we obtain the propagation equation of the density
perturbations in the Bose-Einstein Condensate dark matter fluid as
\begin{equation}  \label{23}
\frac{\partial ^{2}\delta \rho }{\partial t^{2}}=v_{s}^{2}\nabla ^{2}\delta
\rho +4\pi G\rho _{0}\left( 1+\frac{\vec{\Omega}_{0}^{2}}{\pi G\rho _{0}}%
\right) \delta \rho -\frac{\hbar ^{2}}{4m^{2}}\nabla ^{4}\delta \rho .
\end{equation}
To obtain Eq.~(\ref{23}) we have also used the mathematical identity $\nabla
\cdot \left(\delta \vec{v}\times \vec{\Omega }_0\right)=\vec{\Omega }_0\cdot
\left(\nabla \times \delta \vec{v}\right)-\delta \vec{v}\cdot \left(\nabla
\times \vec{\Omega}_0\right)=\vec{\Omega }_0\cdot \left(\nabla \times \delta
\vec{v}\right)$.

We look now for a solution of Eq.~(\ref{23}) of the form $\delta \rho =\sum
A\left( \omega ,\left\vert \vec{k}\right\vert \right) \exp \left[ i\left(
\omega t-\vec{k}\cdot \vec{r}\right) \right] $. Hence after substitution
into the equation of this representation of the density variation we obtain
for each component $\omega $ the following dispersion relation,
\begin{equation}  \label{32}
\omega ^{2}=v_{s}^{2}\vec{k}^{2}-4\pi G\rho _{0}\left( 1+\frac{\vec{\Omega}%
_{0}^{2}}{\pi G\rho _{0}}\right) +\frac{\hbar ^{2}}{4m^{2}}\vec{k}^{4}.
\end{equation}

The Jeans wave number $\left\vert \vec{k}\right\vert _{J}$ corresponds to
the vanishing of the eigenfrequency $\omega $, and it is given as a solution
of the algebraic equation
\begin{equation}
\frac{\hbar ^{2}}{4m^{2}}\vec{k}_{J}^{4}+v_{s}^{2}\vec{k}_{J}^{2}-4\pi G\rho
_{0}\left( 1+\frac{\vec{\Omega}_{0}^{2}}{\pi G\rho _{0}}\right) =0.
\end{equation}%
Hence for $\vec{k}_{J}$ we find
\begin{eqnarray}
\vec{k}_{J}^{2} &=&\frac{2m^{2}}{\hbar ^{2}}v_{s}^{2}\left( \sqrt{1+\frac{%
4\pi G\hbar ^{2}}{m^{2}v_{s}^{4}}\rho _{0}\left( 1+\frac{\vec{\Omega}_{0}^{2}%
}{\pi G\rho _{0}}\right) }-1\right) =  \notag \\
&&\frac{2}{\xi _{h}^{2}}\left( \sqrt{1+\frac{4\pi G\xi _{h}^{2}}{v_{s}^{2}}%
\rho _{0}\left( 1+\frac{\vec{\Omega}_{0}^{2}}{\pi G\rho _{0}}\right) }%
-1\right) ,
\end{eqnarray}%
where we have introduced the healing length of the condensate dark matter
halo as defined by Eq.~(\ref{xi}).

From the dispersion equation, rewritten as
\begin{equation}
\omega ^{2}=\left( \vec{k}^{2}-\vec{k}_{J}^{2}\right) \left[ v_{s}^{2}+\frac{%
\hbar ^{2}}{4m}\left( \vec{k}^{2}+\vec{k}_{J}^{2}\right) \right] ,
\end{equation}%
it follows that for $\left\vert \vec{k}\right\vert <\left\vert \vec{k}%
\right\vert _{J}$, the angular frequency $\omega $ becomes imaginary. This
situation corresponds to an instability of the Bose-Einstein Condensate dark
matter fluid - $\delta \rho $ can either exponentially increase or decrease,
thus leading to a gravitational rarefaction (or condensation) in the dark
matter halo. Hence, for $\left\vert \vec{k}\right\vert <\left\vert \vec{k}%
\right\vert _{J}$, $\omega \propto \pm v_{s}\sqrt{\vec{k}^{2}-\vec{k}_{J}^{2}%
}=i\mathrm{Im}\omega $, and therefore $\delta \rho \propto \exp \left[ \pm
\left\vert \mathrm{Im}\omega \right\vert t\right] $. When the mass of the
Bose-Einstein Condensate dark matter halo is greater than the mass of a
sphere with radius $R_{J}=2\pi /\left\vert \vec{k}\right\vert _{J}$, a
gravitational instability occurs, and the dark matter cloud of particles
would collapse. The critical mass delimitating the two phases is the Jeans
mass $M_{J}=\left( 4\pi /3\right) \left( 2\pi /\left\vert \vec{k}\right\vert
_{J}\right) ^{3}\rho _{0}$. For the Bose Einstein Condensate dark matter
halos in the presence of rotation and vortices, the Jeans radius is given by

\begin{equation}
R_{J}=\frac{\sqrt{2}\pi \hbar }{mv_{s}}\left( \sqrt{1+\frac{4\pi G\hbar ^{2}%
}{m^{2}v_{s}^{4}}\rho _{0}\left( 1+\frac{\vec{\Omega}_{0}^{2}}{\pi G\rho _{0}%
}\right) }-1\right) ^{-1/2},  \label{35}
\end{equation}%
while for the Jeans mass we obtain
\begin{eqnarray}
\hspace{-0.7cm}M_{J} &=&\frac{8\sqrt{2}\pi ^{4}\hbar ^{3}}{3m^{3}v_{s}^{3}}%
\times  \notag  \label{36} \\
\hspace{-0.7cm} &&\left( \sqrt{1+\frac{4\pi G\hbar ^{2}}{m^{2}v_{s}^{4}}\rho
_{0}\left( 1+\frac{\vec{\Omega}_{0}^{2}}{\pi G\rho _{0}}\right) }-1\right)
^{-3/2}\rho _{0}.
\end{eqnarray}

\subsection{Jeans stability of the nonrotating Bose-Einstein Condensate
matter halos}

We consider first the case of the collapse of the very slowly rotating
condensate dark matter halos, having
\begin{eqnarray}
\chi &=&\frac{\vec{\Omega}_{0}^{2}}{\pi G\rho _{0}}=4.774\times 10^{-2}\times
\notag \\
&&\left( \frac{\left\vert \vec{\Omega}\right\vert _{0}}{10^{-16}\;\mathrm{s}%
^{-1}}\right) ^{2}\left( \frac{\rho _{0}}{10^{-24}\;\mathrm{g/cm^{3}}}%
\right) ^{-1}<<1.
\end{eqnarray}

Hence in this approximation the terms containing $\left\vert \vec{\Omega}%
^{2}\right\vert $ can be safely neglected in the expressions of the Jeans
wave number, radius and mass. Therefore for the Jeans wave number of the
slowly rotating/nonrotating Bose Einstein Condensate dark matter halo we
find the expression
\begin{equation}
\left\vert \vec{k}\right\vert _{J}=\frac{\sqrt{2}mv_{s}}{\hbar }\left( \sqrt{%
1+\frac{4\pi G\hbar ^{2}}{m^{2}v_{s}^{4}}\rho _{0}}-1\right)^{1/2} .
\end{equation}

If the condition
\begin{equation}
\frac{4\pi G\hbar^2}{m^2v_s^4}\rho_0=\frac{Gm^4}{4\pi a^2\hbar ^2\rho _0}<<1,
\end{equation}
or, equivalently,
\begin{eqnarray}
\rho _0&>>&\rho _{cr}=\frac{Gm^4}{4\pi a^2\hbar ^2}=4.817\times
10^{-66}\times  \notag \\
&&\left(\frac{m}{\mathrm{meV}}\right)^4\left(\frac{a}{10^{-3}\;\mathrm{fm}}%
\right)^{-2}\;\mathrm{g/cm^3},
\end{eqnarray}
is satisfied, by power expanding the square root we obtain
\begin{eqnarray}
\hspace{-0.7cm}\left|\vec{k}\right|_J&\approx &\sqrt{\frac{4\pi G\rho_0}{%
v_s^2}}=\sqrt{\frac{Gm^3}{\hbar ^2a}}=5.828\times 10^{-23}\times  \notag \\
\hspace{-0.7cm}&&\left(\frac{m}{\mathrm{meV}}\right)^{3/2}\left(\frac{a}{%
10^{-3}\;\mathrm{fm}}\right)^{-1/2}\;\mathrm{cm}^{-1}, \rho _0>>\rho _{cr}.
\end{eqnarray}

In the opposite limit $\left(4\pi G\hbar^2/m^2v_s^4\right)\rho_0>>1$, or
\begin{equation}
\rho _0<<\frac{Gm^4}{4\pi a^2\hbar ^2},
\end{equation}
we obtain
\begin{eqnarray}
\left|\vec{k}\right|_J&\approx&\left(\frac{16\pi Gm^2}{\hbar^2}\rho
_0\right)^{1/4}=1.759\times 10^{-12}\left(\frac{m}{\mathrm{meV}}%
\right)^{1/2}\times  \notag \\
&&\left( \frac{\rho _{0}}{10^{-24}\;\mathrm{g/cm^{3}}}\right) ^{1/4}\;%
\mathrm{cm}^{-1},\;\rho <<\rho _{cr}.
\end{eqnarray}

The effective radius $R_{J}$ of the Bose-Einstein Condensate dark matter
configuration at the onset of the gravitational instability is given by
\begin{equation}
R_{J}=\frac{\sqrt{2}\pi \hbar }{mv_{s}}\left( \sqrt{1+\frac{4\pi G\hbar ^{2}%
}{m^{2}v_{s}^{4}}\rho _{0}}-1\right) ^{-1/2},
\end{equation}%
and it has the limiting values
\begin{eqnarray}\label{Jeans1}
\hspace{-0.5cm}R_J&\approx& \sqrt{\frac{\pi}{G\rho _0}}v_s=\sqrt{\frac{4\pi
^2\hbar ^2a}{Gm^3}}=34.96\times  \notag \\
\hspace{-0.5cm}&&\left(\frac{m}{\mathrm{meV}}\right)^{-3/2}\left(\frac{a}{%
10^{-3}\;\mathrm{fm}}\right)^{1/2}\;\mathrm{kpc}, \rho _0>>\rho_{cr},
\end{eqnarray}
and
\begin{eqnarray}
R_J&\approx& \left(\frac{\pi ^3\hbar ^2}{Gm^2\rho _0}\right)^{1/4}=3.57%
\times 10^{12}\times \left(\frac{m}{\mathrm{meV}}\right)^{-1/2}\times  \notag
\\
&&\left( \frac{\rho _{0}}{10^{-24}\;\mathrm{g/cm^{3}}}\right) ^{-1/4}\;%
\mathrm{cm}, \rho _0<<\rho _{cr},
\end{eqnarray}
respectively. The Jeans mass of the static Bose-Einstein Condensate dark
matter halo is given by
\begin{equation}
M_{J}=\frac{16\pi ^{4}\hbar ^{3}\rho _{0}}{3\sqrt{2}m^{3}v_{s}^{3}}\left(
\sqrt{1+\frac{4\pi G\hbar ^{2}}{m^{2}v_{s}^{4}}}-1\right) ^{-3/2}.
\end{equation}

For $\rho _{0}>>\rho _{cr}$, for the Jeans mass we obtain the approximate
expression
\begin{eqnarray}  \label{34}
M_J&\approx &\frac{32\pi^4}{3}\left(\frac{\hbar ^2a}{Gm^3}\right)^{3/2}\rho
_0=2.623\times 10^{12} \times  \notag \\
&&\left(\frac{a}{10^{-3}\;\mathrm{fm}}\right)^{3/2} \times\left(\frac{m}{%
\mathrm{meV}}\right)^{-9/2}\times  \notag \\
&&\left(\frac{\rho _0}{10^{-24}\;\mathrm{g/cm^3}}\right)M_{\odot}, \;\rho
>>\rho _{cr}.
\end{eqnarray}
For $\rho _0<<\rho _{cr}$ we find
\begin{eqnarray}
\hspace{-0.6cm}M_J&\approx& \frac{32\pi ^4}{3}\left(\frac{\hbar ^2}{16\pi
Gm^2}\right)^{3/4}\rho _0^{1/4}=1.906\times 10^{14}\times  \notag \\
\hspace{-0.6cm}&&\left(\frac{m}{\mathrm{meV}}\right)^{-3/2}\left(\frac{\rho
_0}{10^{-24}\;\mathrm{g/cm^3}}\right)^{1/4}\;\mathrm{g}, \rho_0 <<\rho _{cr}.
\end{eqnarray}

The maximum mass of the high density astrophysical objects, like white
dwarfs and neutron stars, described by polytropic equations of state, was
derived by Landau and Chandrasekhar, and it can be approximated by the
Chandrasekhar mass \cite{Chandb}, \cite{Shap},
\begin{equation}  \label{chandra}
M_{Ch}\approx \left( \frac{\hbar c}{G}\right) ^{3/2}\frac{1}{m_B^2},
\end{equation}
where by $m_{B}$ we have denoted the mass of the particles that give the
most important contribution to the total stellar mass (protons and neutrons
in the case of the white dwarfs and neutron stars, respectively)~ \cite%
{Chandb}, \cite{Shap}. Hence, the maximum mass of a degenerate star is fully
determined by fundamental physical constants only, with the possible
exception of some numerical factors that depend on the chemical composition
of the star.

Interestingly enough, the Jeans mass for Bose-Einstein Condensate dark
matter halos given by Eq.~(\ref{34}) can be also written in a form similar
to the Chandrasekhar maximum mass, if we associate to the dark matter
particles an effective mass $m_{eff}$ defined as
\begin{equation}
m_{eff}=\frac{1}{\sqrt{\rho_0}}\left(\frac{cm^3}{\hbar a}\right)^{3/4},
\end{equation}
so that
\begin{equation}
M_{J}=\frac{8\sqrt{2}\pi ^4}{3}\left(\frac{\hbar c}{G}%
\right)^{3/2}m_{eff}^{-2}.
\end{equation}
For the adopted values of the physical parameters of the dark matter halos
the numerical value of the effective mass is of the order of $m_{eff}\approx
4.52\times 10^{-29}$ g.

On the other hand, one could also ask the question if Bose-Einstein
Condensate dark matter could form astrophysical objects and structures with
the limiting mass given by the Chandrasekhar mass as given by Eq.~(\ref%
{chandra}). For the adopted mass of the dark matter particle $m=1 \;\mathrm{%
meV}=1.78\times 10^{-36}$ g, the Chandrasekhar limit gives a mass of the
order of $M_{Ch}=1.62\times 10^{34}M_{\odot}=3.25\times 10^{57}$ g, which
exceeds by three orders of magnitude the total mass of the ordinary matter
in the Universe, $4.5\times 10^{54}$ g \cite{Planck}. Therefore, it follows
that the Chandrasekhar mass limit cannot lead to realistic descriptions of
the physical properties of condensate dark matter clouds composed of
particles having masses of the order of the mass values assumed for
condensate dark matter particles.

\subsection{The effects of the rotation}

Generally, in the slow rotation limit when the factor $\vec{\Omega}%
_{0}^{2}/\pi G\rho _{0}$ is still small, or of the order of unity, the
rotation modifies the Jeans parameters (Jeans wavelength, radius and mass,
respectively), by a factor of $1+\chi $. Hence, if, for example, the
condition
\begin{equation}
\frac{4\pi G\hbar ^{2}}{m^{2}v_{s}^{4}}\rho _{0}\left( 1+\frac{\vec{\Omega}%
_{0}^{2}}{\pi G\rho _{0}}\right) <<1,
\end{equation}%
holds, then we immediately obtain

\begin{equation}
\left\vert \vec{k}\right\vert _{J}\approx \sqrt{\frac{Gm^{3}}{\hbar ^{2}a}}%
\sqrt{1+\frac{\vec{\Omega}_{0}^{2}}{\pi G\rho _{0}}},
\end{equation}

\begin{equation}
R_{J}\approx \sqrt{\frac{4\pi ^{2}\hbar ^{2}a}{Gm^{3}}}\left( 1+\frac{\vec{%
\Omega}_{0}^{2}}{\pi G\rho _{0}}\right) ^{-1/2},
\end{equation}%
and

\begin{equation}
M_{J}\approx \frac{32\pi ^{4}}{3}\left( \frac{\hbar ^{2}a}{Gm^{3}}\right)
^{3/2}\left( 1+\frac{\vec{\Omega}_{0}^{2}}{\pi G\rho _{0}}\right)
^{-3/2}\rho _{0},
\end{equation}%
respectively. However, with the increase of the initial rotational velocity
of the Bose-Einstein condensed dark matter halo, the effects of the rotation
become more and more important. For a rotational angular velocity of $%
\left\vert \vec{\Omega}_{0}\right\vert =10^{-14}$ s$^{-1}$, we obtain $\chi =%
\vec{\Omega}_{0}^{2}/\pi G\rho _{0}=477.47$, while for $\left\vert \vec{%
\Omega}_{0}\right\vert =10^{-12}$ s$^{-1}$ we have $\chi =\allowbreak
4.\,\allowbreak 77\times 10^{6}$. Hence for the astrophysical and physical
description of dark matter halos rotating with such initial angular
velocities the effect of the rotation must be taken into account. Hence for
rapidly rotating dark matter condensed clouds the Jeans wavelength becomes

\begin{equation}
\left\vert \vec{k}_{J}\right\vert \approx \frac{\sqrt{2}m}{\hbar }%
v_{s}\left( \sqrt{1+\frac{4\hbar ^{2}\vec{\Omega}_{0}^{2}}{m^{2}v_{s}^{4}}}%
-1\right) ^{1/2},
\end{equation}%
or, equivalently,

\begin{equation}
\left\vert \vec{k}_{J}\right\vert \approx \sqrt{\frac{8\pi a}{m}\rho _{0}}%
\left( \sqrt{1+\frac{m^{4}\vec{\Omega}_{0}^{2}}{4\pi ^{2}\hbar ^{2}a^{2}\rho
_{0}^{2}}}-1\right) ^{1/2}.
\end{equation}

For the adopted values of the physical and astrophysical quantities the
parameter $m^{4}\vec{\Omega}_{0}^{2}/4\pi ^{2}\hbar ^{2}a^{2}\rho _{0}^{2}$
has generally small values, much smaller than one. Therefore we ca expand
the square root into power series, and in the first approximation we obtain
\begin{eqnarray}
\hspace{-0.3cm}&&\left\vert \vec{k}_{J}\right\vert \approx \sqrt{\frac{2}{%
\pi }}\frac{m^{3/2}}{\hbar \sqrt{a\rho _{0}}}\left\vert \vec{\Omega}%
_{0}\right\vert =1.801\times 10^{-19}\left( \frac{m}{\mathrm{meV}}\right)
^{3/2} \times  \notag \\
\hspace{-0.3cm}&&\left( \frac{a}{10^{-3}\;\mathrm{fm}}\right) ^{-1/2}\left(
\frac{\rho _{0}}{10^{-24}\;\mathrm{g/cm^{3}}}\right) ^{-1/2}\frac{\left\vert
\vec{\Omega}_{0}\right\vert }{10^{-12}s^{-1}}\;\mathrm{cm}^{-1}.  \notag \\
\end{eqnarray}

Hence the Jeans radius and the Jeans mass of rapidly rotating Bose-Einstein
condensate dark matter halos can be obtained as
\begin{eqnarray}
\hspace{-0.7cm}&&R_{J} \approx \frac{\sqrt{2}\pi ^{3/2}\hbar \sqrt{a\rho _{0}}}{%
m^{3/2}\left\vert \vec{\Omega}_{0}\right\vert }= 1. 132\times 10^{-2}\times
\left( \frac{m}{\mathrm{meV}}\right) ^{-3/2}\times  \notag \\
\hspace{-0.7cm}&&\left( \frac{a}{10^{-3}\;\mathrm{fm}}\right) ^{1/2}\left( \frac{\rho _{0}}{%
10^{-24}\;\mathrm{g/cm^{3}}}\right) ^{1/2}\left( \frac{\left\vert \vec{\Omega%
}_{0}\right\vert }{10^{-12}s^{-1}}\right) ^{-1}\;\mathrm{kpc},  \nonumber\\
\end{eqnarray}
and
\begin{eqnarray}
M_{J} &\approx& \frac{8\sqrt{2}\pi ^{11/2}}{3}\frac{\hbar ^{3}\left( a\rho
_{0}^{5/3}\right) ^{\frac{3}{2}}}{m^{9/2}\left\vert \vec{\Omega}%
_{0}\right\vert ^{3}}= 88. 752\times \left( \frac{m}{\mathrm{meV}}\right)
^{-9/2}\times  \notag \\
&& \left( \frac{a}{10^{-3}\;\mathrm{fm}}\right) ^{3/2}\left( \frac{\rho _{0}%
}{10^{-24}\;\mathrm{g/cm^{3}}}\right) ^{3/2}\times  \notag \\
&&\left( \frac{\left\vert \vec{\Omega}_{0}\right\vert }{10^{-12}s^{-1}}%
\right) ^{-3}\left(\frac{\rho _0}{10^{-24}\;\mathrm{g/cm^3}}\right)M_{\odot
},
\end{eqnarray}
respectively. Interestingly enough, in the case of the rotation dominated
Bose-Einstein Condensate dark matter halos, their physical and astrophysical
properties are independent of the gravitational constant $G$, and they are
fully determined by the fundamental quantum parameters $m$ and $a$,
characterizing the condensate, as well as by the angular velocity of the
halo $\left\vert \vec{\Omega}_{0}\right\vert$.

\subsection{The Thomas-Fermi approximation}

Finally, as a last step in our general analysis of the properties of the
Bose-Einstein Condensate dark matter halos we introduce the Thomas-Fermi
approximation for the hydrodynamical description of the condensate. In order
to perform our analysis, we decompose the total energy, $E_{tot}$, into
kinetic, interaction, quantum and gravitational contributions \cite{Dalfovo,Pita,Pethick,ZNG},
\begin{equation}
E_{tot}=E_{kin}+E_{int}+E_q+E_{grav},
\end{equation}
where
\begin{equation}
E_{kin}=\frac{\hbar ^2}{2m}\int{\left[\sqrt{\rho \left(\vec{r},t\right)}\vec{%
v}\left(\vec{r},t\right)\right]^2d^3\vec{r}},
\end{equation}
\begin{equation}
E_{int}=U_0\int{\left[\rho \left(\vec{r},t\right)\right]^2d^3\vec{r}},
\end{equation}
\begin{equation}
E_q=\frac{\hbar ^2}{2m}\int{\left[\nabla \sqrt{\rho \left(\vec{r},t\right)}%
\right]^2d^3\vec{r}},
\end{equation}
\begin{equation}
E_{grav}=\int{\rho \left(\vec{r},t\right)\phi \left(\vec{r},t\right)d^3\vec{r%
}.}
\end{equation}

The kinetic energy is determined by the velocity field of the dark matter
fluid. The interaction energy is related to the interaction of the particles
in the condensate. The quantum energy is determined by square of the
gradient of the square root of the density of the condensate, while the
gravitational energy has its origin from the external potential of the
field, acting as a trapping potential.

In the Thomas-Fermi approximation, we neglect in the total anergy balance
the kinetic energy term $-\left( \hbar ^{2}/2m\right) \Delta $ of the
condensate particles. To obtain the conditions of the validity of the
Thomas-Fermi approximation we consider a system composed of $N$ bosons in a
volume $V$, extended over a radius $R$, in the presence of a confining
gravitational potential. The total mass of the system is denoted by $M$.
Moreover, we assume that all bosons are in the same quantum state. The total
energy $E_{tot}$ of the system is given by $E=E_{kin}+E_{int}+E_{grav}$. For
a single particle the kinetic energy can be approximated as $\hbar ^2/2mR^2$
\cite{Pethick}, and hence the total kinetic energy of the system can be obtained
as $E_{kin}=N\hbar ^2/2mR^2$. The interaction energy is given by $%
E_{int}=(1/2) \left(N^2/V\right)mg$ \cite{Pethick}, while the gravitational
potential energy can be approximated as $E_{grav}=-\alpha GM^2/R$, where $\alpha $ is a constant. For the $n=1$ polytropic equation of state $\alpha =3/4$ \cite{Horedt}. Therefore for the
total energy of the bosonic system in a gravitational field we obtain the
expression
\begin{equation}
E_{tot}=N\frac{\hbar ^2}{2mR^2}+\frac{3}{2}N^2\frac{\hbar ^2a}{mR^3}-\frac{\alpha %
GM^2}{R}.
\end{equation}

If the condition
\begin{equation}
\frac{3Na}{R}>>1,
\end{equation}
is satisfied, then the interaction energy is much larger than the kinetic
energy, $E_{int}>>E_{kin}$. In order to check the validity of the condition
in an astrophysical setting, we consider a galactic condensate of mass $%
M=10^{10}M_{\odot}=2\times 10^{43}$ g. Then number of the dark matter
particle is of the order of $N=M/m=1.12\times 10^{79}$ particles. If the
radius of the condensate is $R=10$ kpc = $3\times 10^{22}$ cm, then the
quantity $3Na/R=1.12\times 10^{41}$, a number that is obviously much greater
than one, $3Na/R>>1$. Hence in galactic Bose-Einstein Condensate dark matter
halos the contribution to the total energy of the kinetic energy of the
particles can be safely neglected when compared to the interaction energy of
the bosonic fluid. Therefore, in the case of bosonic systems consisting of a
very large number of particles, the Thomas-Fermi approximation gives an
excellent description of the physical and astrophysical properties of the
condensate, corresponding to a precision level that can be recognized as
\textit{exact}.

On the other hand, the physical length scales of the order of $R\approx
\sqrt{m/4\pi \rho a}$, where $\rho $ is the mean dark matter density, the
Thomas-Fermi approximation is not valid anymore. For $\rho =10^{-24}\;%
\mathrm{g/cm^3}$, we obtain $R\approx 377.65$ cm, a length scale that is
insignificant from the point of view of the galactic dark matter
distribution. For the gravitational energy of the Bose-Einstein Condensate
dark matter halo we obtain the value $E_{grav}=8.89\times 10^{56}$ ergs, why
for the considered numerical values of the parameters of the dark matter
halos the interaction energy has the value $E_{int}=1.35\times 10^{56}$
ergs. Hence in gravitationally trapped galactic size condensates the
interaction energy is of the same order of magnitude as the gravitational
energy of the bosons. However, it is important to point out that $%
E_{grav}>E_{int}$, a relation that shows that the dark matter halo is
trapped by its gravitational energy.

Hence in the Thomas-Fermi approximation the total energy of the condensate dark matter halo is given by
\be
E_{tot}=\frac{3}{2}M^2\frac{\hbar ^2a}{m^3}\frac{1}{R^3}-\frac{3}{4}\frac{GM^2}{R}.
\ee

The equilibrium radius corresponding to the minimum of the total energy is given by the solution of the algebraic equation
\be
\frac{dE_{tot}}{dR}=-\frac{9}{2}M^2\frac{\hbar ^2a}{m^3}\frac{1}{R^4}+\frac{3}{4}\frac{GM^2}{R^2}=0,
\ee
given by
\be
R=\sqrt{\frac{6\hbar ^2a}{Gm^3}},
\ee
which, except some numerical factors, coincides with Eq.~(\ref{Jeans1}), giving the Jeans radius of the condensate dark matter halo.

Therefore in our subsequent analysis of the gravitational collapse of
Bose-Einstein Condensate dark matter halos we will neglect, with a very good
approximation, the kinetic energy term in the Gross-Pitaevskii Eq.~(\ref{sch}%
), and, consequently, the quantum force, derived from the quantum potential,
in the hydrodynamic description of the dynamics of the condensate. In the
Thomas-Fermi approximation the chemical potential of the static condensate
is given by
\begin{equation}  \label{8b}
\mu =m\phi \left( \vec{r}\right) +U_0 \rho \left( \vec{r}\right) .
\end{equation}

\section{Time evolution of Bose-Einstein Condensate dark matter halos}\label{sect3}

We consider now a Bose-Einstein Condensate dark matter halo falling inward
from an initial rest state. The halo is only under the effects of its own
self-gravity. The basic equations describing the dynamical evolution of the
condensate dark matter halo are then the continuity equation, the Euler
equation, and the Poisson equation, which in spherical symmetry are given by
\begin{equation}  \label{s1}
\frac{\partial \rho}{\partial t}+\frac{1}{r^2}\frac{\partial }{\partial r}%
\left(r^2\rho v\right)=0,
\end{equation}
\begin{equation}  \label{s2}
\frac{\partial v}{\partial t}+v\frac{\partial v}{\partial r}=-2u_0\frac{%
\partial \rho}{\partial r}-\frac{\partial \phi}{\partial r}+\frac{2}{3}%
\Omega ^2r,
\end{equation}
\begin{equation}  \label{s3}
\frac{1}{r^2}\frac{\partial }{\partial r}\left(r^2\frac{\partial \phi}{%
\partial r}\right)=4\pi G\rho,
\end{equation}
where in the Euler equation we have used the polytropic equation of state $%
p=u_0\rho ^2$ of the condensate. The last term in Eq.~(\ref{s2}) takes into
account the effects of the rotation, and it has been obtained as follows. If
$\theta $ is the colatitude of a given point in the condensate with respect
to the rotation axis, the centrifugal acceleration of the former is $\Omega
^2 r\sin \theta$. Its radial component is $\Omega ^2 r\sin ^2\theta$. In
order to preserve the spherical symmetry of the dark matter halo we will
neglect all the other components of the centrifugal acceleration, and take
into account the mean value of $\Omega ^2 r\sin ^2\theta$ on a sphere of
radius $r$. Consequently we obtain for the centrifugal acceleration the
expression $2\Omega ^2r/3$ \cite{Kipp}.

The Poisson equation (\ref{s3}) can be immediately integrated to give
\begin{equation}  \label{grad}
\frac{\partial \phi}{\partial r}=\frac{GM\left(r,t\right)}{r^2},
\end{equation}
where $M(r,t)$, the total mass of the dark matter halo within radius $r$, is
defined as
\begin{equation}
M(r,t)=4\pi \int_0^r{r^{\prime }{^2}\rho \left(r^{\prime
},t\right)dr^{\prime }}.
\end{equation}

We multiply now the continuity equation (\ref{s1}) by $4\pi r^2$. After
using the relation $4\pi r^2\rho =\partial M/\partial r$, and integrating
with respect to $r$, we obtain
\begin{equation}  \label{s4}
\frac{\partial M(r,t)}{\partial t}+v(r,t)\frac{\partial M(r,t)}{\partial r}%
=0.
\end{equation}

By using the expression of the gradient of the gravitational potential as
given by Eq.~(\ref{grad}), the Euler equation (\ref{s2}) becomes
\begin{eqnarray}  \label{s5}
\hspace{-0.4cm}\frac{\partial v(r,t)}{\partial t}+v(r,t)\frac{\partial v(r,t)}{\partial r}%
&=&-2u_0\frac{\partial \rho(r,t)}{\partial r}-  \notag \\
\hspace{-0.4cm}&&\frac{GM(r,t)}{r^2}+\frac{2}{3}\Omega ^2(t)r.
\end{eqnarray}
Equations (\ref{s4}) and (\ref{s5}) represent a system of two coupled
nonlinear partial differential equations whose solutions for $v$ and $M$
fully determine the dynamical evolution of the collapsing Bose-Einstein
Condensate dark matter halo. This system of equations can be reduced to a
single equation describing the dynamical evolution of the collapse. From
Eq.~(\ref{s4}) we obtain
\begin{equation}  \label{vel}
v(r,t)=-\frac{\dot{M}(r,t)}{M^{\prime }(r,t)}.
\end{equation}
Substituting the above expression of the velocity into Eq.~(\ref{s5}), and
using the definition of the density in terms of the mass, we obtain the mass
equation describing the time evolution of the Bose-Einstein Condensate dark
matter halo as
\begin{eqnarray}  \label{massa}
\hspace{-0.9cm} &&-\frac{\ddot{M}(r,t)}{M^{\prime }(r,t)}+\frac{2\dot{M}(r,t)%
\dot{M}^{\prime }(r,t)}{M^{\prime 2}(r,t)}-\frac{\dot{M}^2(r,t)M^{\prime
\prime }(r,t)}{M^{\prime 3}(r,t)}=  \notag \\
\hspace{-0.9cm}&& \frac{u_0}{\pi r^3}M^{\prime }(r,t)-\frac{u_0}{2\pi r^2}%
M^{\prime \prime }(r,t)- \frac{GM(r,t)}{r^2}+\frac{2}{3}\Omega ^2(t)r.
\end{eqnarray}

\subsection{The stationary solution}

We consider first the case of the rotating stationary (static) Bose-Einstein
Condensate dark matter halos. The stationary case corresponds to $\vec{v}%
\equiv 0$, $M=M_0(r)$, and $\Omega =\mathrm{constant}$, respectively. In
this situation the mass evolution equation Eq.~(\ref{massa}) takes the form
\begin{equation}  \label{79}
\frac{d^2M_0(r)}{dr^2}-\frac{2}{r}\frac{dM_0(r)}{dr}+a_0M_0(r)-b_0r^3=0,
\end{equation}
where we have denoted
\begin{equation}
a_0=\frac{2\pi G}{u_0}=\frac{Gm^3}{\hbar ^2 a},
\end{equation}
and
\begin{equation}
b_0=\frac{4\pi \Omega ^2}{3u_0}=\frac{2\Omega ^2m^3}{3\hbar ^2a},
\end{equation}
respectively. Eq.~(\ref{79}) has the general solution
\begin{eqnarray}
M_0(r)&=&\sqrt{\frac{2}{\pi }}a_0^{-3/4} \Bigg[\left(c_1-\sqrt{a_0} c_2
r\right) \sin \left(\sqrt{a_0} r\right)-  \notag \\
&& \left(\sqrt{a_0} c_1 r+c_2\right) \cos \left(\sqrt{a_0} r\right)\Bigg]+
\frac{b_0 r^3}{a_0},
\end{eqnarray}
where $c_1$ and $c_2$ are arbitrary constants of integration. The boundary
condition
\begin{equation}
M_0(0)=-\sqrt{\frac{2}{\pi}}a_0^{-3/4}c_2=0,
\end{equation}
gives $c_2$=0, and therefore for the mass of the stationary Bose-Einstein
Condensate dark matter halo we obtain
\begin{eqnarray}
M_0(r)&=&\sqrt{\frac{2}{\pi }}a_0^{-3/4} c_1 \left[\sin \left(\sqrt{a_0}
r\right)-\sqrt{a_0} r \cos \left(\sqrt{a_0} r\right)\right]+  \notag \\
&& \frac{b_0 r^3}{a_0}.
\end{eqnarray}
For the density distribution of the rotating condensate we obtain
\begin{equation}
\rho _0(r)=\frac{3 b_0}{4 \pi a_0}+\frac{\sqrt[4]{a_0} c_1 \sin \left(\sqrt{%
a_0} r\right)}{2 \sqrt{2} \pi ^{3/2} r}
\end{equation}

The condition
\begin{equation}
\rho_0(0)=\rho _c=\frac{a_0^{3/4} c_1}{2 \sqrt{2} \pi ^{3/2}}+\frac{3 b_0}{4
\pi a_0},
\end{equation}
where $\rho _c$ is the central density of the condensate, gives for the
integration constant $c_1$ the expression
\begin{equation}
c_1=\sqrt{\frac{\pi }{2}}a_0^{-7/4} \left(4 \pi a_0 \rho_c-3 b_0\right).
\end{equation}

Hence we finally obtain for the density and mass distribution of the
rotating Bose-Einstein Condensate dark matter halos the expressions
\begin{equation}
\rho _0(r)=\frac{\Omega ^2}{2\pi G}+\rho _c\left(1-\frac{\Omega ^2}{2\pi
G\rho _c}\right)\frac{\sin \left(\sqrt{a_0}r\right)}{\sqrt{a_0}r},
\end{equation}
and
\begin{eqnarray}\label{93}
M_0(r)&=&\frac{2\Omega ^2}{3G}r^3+4\pi a_0^{-3/2}\rho _c\left(1-\frac{\Omega
^2}{2\pi G\rho _c}\right)\times  \notag \\
&&\left[\sin \left(\sqrt{a_0} r\right)-\sqrt{a_0} r \cos \left(\sqrt{a_0}
r\right)\right],
\end{eqnarray}
respectively. The radius $R_0$ of the stationary condensate is determined
from the condition $\rho \left(R_0\right)=0$, and it can be obtained as a
solution of the equation
\begin{equation}
\sin \left(\sqrt{a_0}R_0\right)=\sqrt{a_0}R_0\frac{\Omega ^2}{2\pi \rho _c}%
\left(\frac{\Omega ^2}{2\pi G\rho _c}-1\right)^{-1}=-\lambda \sqrt{a_0}R_0,
\end{equation}
where we have denoted
\begin{equation}
\lambda =\frac{\Omega ^2}{2\pi G\rho _c}\left(1-\frac{\Omega ^2}{2\pi G \rho
_c}\right)^{-1}.
\end{equation}
By taking into account that the left hand side of the above equation can be
expanded in Taylor series as
\begin{equation}
\sin \left(\sqrt{a_0}R_0\right)=\sqrt{a_0} R_0-\frac{1}{6} a_0^{3/2} R_0^3+%
\frac{1}{120} a_0^{5/2} R_0^5+O\left(R^6\right),
\end{equation}
we obtain for the determination of the radius of the rotating condensate
dark matter halo the algebraic equation
\begin{equation}
\sqrt{a_0}R_0\left(1+\lambda -\frac{1}{6}a_0R_0^2+\frac{1}{120}%
a_0^2R_0^4+...\right)=0.
\end{equation}

In the first order of approximation we obtain for the stationary dark matter
halo the approximate expression
\begin{equation}
R_0\approx \sqrt{\frac{6(1+\lambda)}{a_0}}=\sqrt{\frac{6}{\left(1-\Omega
^2/2\pi G\rho _c\right)}\frac{\hbar ^2a}{Gm^3}}.
\end{equation}
In the case of the static condensate with $\Omega \equiv 0$, the radius $R_S$
of the dark matter halo is obtained exactly as
\bea
\hspace{-0.5cm}R_S&=&\pi \sqrt{\frac{\hbar ^2a}{Gm^3}}=\pi \sqrt{a_0}=\nonumber\\
\hspace{-0.5cm}&&17.96\times \left(\frac{a}{10^{-3}\;%
\mathrm{fm}}\right)^{1/2}\times \left(\frac{m}{\mathrm{meV}}\right)^{-3/2}\;
\mathrm{kpc}.
\eea
Then the radius of the rotating, stationary condensate dark matter halo can
be also written in the form
\begin{equation}
R_0\approx \sqrt{\frac{6}{\pi ^2\left(1-\Omega ^2/2\pi G \rho _c\right)}}R_S.
\end{equation}

The total mass of the rotating condensate dark matter halo is obtained as
\begin{eqnarray}
M_0\left(R_0\right)&=&\frac{2\Omega ^2}{3G}R_0^3-\frac{4\pi\hbar ^2a}{Gm^3}%
\rho _cR_0\left(1-\frac{\Omega ^2}{2\pi G\rho _c}\right)\times  \notag \\
&&\left[\lambda +\sqrt{1-\lambda ^2a_0R_0^2}\right].
\end{eqnarray}
In the case of the static condensate $\sqrt{a_0}R_S=\pi$, and for the total
mass $M_S$ of the nonrotating dark matter halo we find
\begin{equation}
M_S=4\pi a_0^{-3/2}=4\pi \left(\frac{\hbar ^2a}{Gm^3}\right)^{3/2}\rho _c=%
\frac{4}{\pi}\rho _cR_S^3,
\end{equation}
or
\begin{eqnarray}
M_S&=&3.17\times 10^{10}\times \left(\frac{a}{10^{-3}\;\mathrm{fm}}%
\right)^{3/2}\times \left(\frac{m}{\mathrm{meV}}\right)^{-9/2}\times  \notag
\\
&&\left(\frac{\rho _c}{10^{-24}\;\mathrm{g/cm^3}}\right)M_{\odot}.
\end{eqnarray}
In the first order approximation the mass of the rotating condensate can be
expressed in terms of the static radius as
\begin{eqnarray}
\hspace{-0.6cm}M_0\left(R_0\right)&\approx &\frac{2\Omega ^2}{3G}R_0^3-\frac{%
4\sqrt{6}}{\pi ^2}\rho _cR_S^3\left(1-\frac{\Omega ^2}{2\pi G\rho _c}%
\right)^{1/2}\times  \notag \\
\hspace{-0.6cm}&&\left[\lambda +\sqrt{1-\left(\frac{\Omega ^2}{2\pi G\rho _c}%
\right)^2\left(1-\frac{\Omega ^2}{2\pi G\rho _c}\right)^{-3}}\right].  \notag
\\
\end{eqnarray}

\subsection{Collapsing Bose-Einstein Condensate dark matter halos}

Obtaining some analytical or semianalytical solutions of the equations
describing the time evolution of Bose-Einstein Condensate dark matter halos
would allow to construct a clearer picture of the relationship between the
input physics and the behavior of the system, also making possible to
understand astrophysical effects that are more difficult to understand with
the extensive use of numerical methods for solving the hydrodynamical
evolution equations. One of the powerful mathematical approaches for the
study of the collapse processes in astrophysical phenomena is based on the
idea of self-similarity, which consists in the rescaling of the radial
coordinate $r$ as $r\rightarrow \xi =r/\alpha(t)$, where $\alpha (t)$ is a
time only dependent function, and to assume that all the other physical
parameters can be expressed as functions of $\xi $ and of some function of
time, so that an arbitrary physical quantity $\Phi (r,t)$ can be written as $%
\Phi (r,t)=\beta (t)\phi (\xi)$. An interesting consequence of the
self-similar (homologous) evolution is that the initial density and mass
profiles of the collapsing systems do not change. From a mathematical point
of view after introducing the appropriately chosen self-similarity
transformations, the system of nonlinear partial differential equations (\ref%
{s4}) and (\ref{s5}) can be reduced in many cases to a system of ordinary
nonlinear differential equations. Some astrophysically relevant self-similar
solutions describe the gravitational collapse of isothermal spheres or of
polytropic spheres. It would be then interesting to consider self-similar
solutions of the hydrodynamic equations describing the time evolution of
Bose-Einstein Condensate Dark matter halos. Unfortunately, the hydrodynamic
equations describing the evolution of a polytropic gas with equation of
state $p\propto \rho ^2$ do not admit self-similar or homologous solutions
(for a proof of this results see Appendix~\ref{appa}. The main reason is
that after introducing the self-similar variables in the standard way, the
explicit time dependence of the equations cannot be eliminated for any
choice of the time similarity functions.

\subsubsection{The evolution equations}

However, the dynamic evolution equation of the mass of the Bose-Einstein
Condensate dark matter halos, described by a polytropic equation of state
with $n=1$, and given by Eq.~(\ref{massa}), does admit a separable,
semi-analytical solution, which can be obtained under the assumption that
the mass of the condensate can be represented as
\begin{equation}
M(r,t)=f(t)m(r),
\end{equation}
where $f(t)$ and $m(r)$ are functions depending on the time, and the radial
coordinate $r$ only. Then the velocity of the dark matter halo can be
immediately obtained from Eq.~(\ref{vel}) as
\begin{equation}
v(r,t)=-\frac{\dot{f}(t)}{f(t)}\frac{m(r)}{m^{\prime }(r)}.
\end{equation}
For the acceleration of the fluid we find
\begin{eqnarray}
\hspace{-0.5cm}&&\frac{dv}{dt}=\left(-\frac{\ddot{f}(t)}{f(t)}+\frac{\dot{f}^2(t)}{f^2(t)}%
\right)\frac{m(r)}{m^{\prime }(r)}+\frac{\dot{f}^2(t)}{f^2(t)}\frac{m(r)}{%
m^{\prime }(r)}\frac{d}{dr}\frac{m(r)}{m^{\prime }(r)}=  \notag \\
\hspace{-0.5cm}&&\left(-\frac{\ddot{f}(t)}{f(t)}+\frac{\dot{f}^2(t)}{f^2(t)}\right)\frac{%
m(r)}{m^{\prime }(r)}+\frac{1}{2}\frac{\dot{f}^2(t)}{f^2(t)}\frac{d}{dr}%
\left[\frac{m(r)}{m^{\prime }(r)}\right]^2.
\end{eqnarray}

Then the mass equation (\ref{massa}) can be written as
\begin{eqnarray}  \label{89}
\hspace{-0.6cm}&&\left(-\frac{\ddot{f}(t)}{f(t)}+\frac{\dot{f}^2(t)}{f^2(t)}%
\right)\frac{m(r)}{m^{\prime }(r)}+\frac{1}{2}\frac{\dot{f}^2(t)}{f^2(t)}%
\frac{d}{dr}\left[\frac{m(r)}{m^{\prime }(r)}\right]^2=  \notag \\
\hspace{-0.6cm}&&f(t)\left[\frac{u_0}{\pi r^3}m^{\prime }(r)-\frac{u_0}{2\pi
r^2}m^{\prime \prime }(r)- \frac{Gm(r)}{r^2}\right]+\frac{2}{3}\Omega ^2(t)r.
\notag \\
\end{eqnarray}

The explicit time dependence in Eq.~(\ref{89}) can be removed, and the
variables can be separated, if the function $f(t)$ and the angular velocity $%
\Omega (t)$ satisfy the conditions
\begin{equation}
-\frac{\ddot{f}(t)}{f(t)}+\frac{\dot{f}^{2}(t)}{f^{2}(t)}=\alpha _{0}f(t),
\label{90}
\end{equation}%
\begin{equation}
\frac{\dot{f}^{2}(t)}{f^{2}(t)}=\frac{4}{\beta _{0}^{2}}f(t),  \label{91}
\end{equation}%
\begin{equation}
\Omega ^{2}(t)=\omega _{0}^{2}f(t),  \label{92}
\end{equation}%
where $\alpha _{0}$, $\beta _{0}$ and $\omega _{0}$ are \textit{dimensional}
constants. Eq.~(\ref{91}) can be integrated immediately to give
\begin{equation}
f(t)=\frac{\beta _{0}^{2}}{\left( t_{0}\pm t\right) ^{2}},
\end{equation}%
where $t_{0}$ is an arbitrary constant of integration. Then Eq.~(\ref{90})
reduces to
\begin{equation}
-\frac{\ddot{f}(t)}{f(t)}+\frac{\dot{f}^{2}(t)}{f^{2}(t)}-\alpha _{0}f(t)=-%
\frac{2+\alpha _{0}\beta _{0}^{2}}{\left( t_{0}\pm t\right) ^{2}}=0,
\end{equation}%
giving $\alpha _{0}=-2/\beta _{0}^{2}$. Hence, the time dependence in the
mass evolution equation cancels out, and for the radial mass distribution of
the time evolving dark matter halo we obtain the equation
\begin{eqnarray}
&&\left[ \frac{u_{0}\beta _{0}^{2}}{4\pi r^{2}}m^{\prime \prime }(r)-\frac{%
u_{0}\beta _{0}^{2}}{2\pi r^{3}}m^{\prime }(r)+\frac{G\beta _{0}^{2}m(r)}{%
2r^{2}}\right] -\frac{1}{3}\omega _{0}^{2}\beta _{0}^{2}r=  \notag
\label{95} \\
&&\frac{m(r)}{m^{\prime }(r)}-\frac{d}{dr}\left[ \frac{m(r)}{m^{\prime }(r)}%
\right] ^{2}.
\end{eqnarray}%
In the above equation $\beta _{0}$ has the physical unit of second, while $%
\omega _{0}$ has the units $1/s$. The function $f(t)$ is dimensionless.
Hence once the solution of Eq.~(\ref{95}) is known, the evolution of the
physical parameters of the collapsing/expanding Bose-Einstein dark matter
halos are given by
\begin{eqnarray}
M(r,t) &=&\frac{\beta _{0}^{2}}{\left( t_{0}\pm t\right) ^{2}}m(r),\rho
(r,t)=\frac{\beta _{0}^{2}}{4\pi r^{2}\left( t_{0}\pm t\right) ^{2}}%
m^{\prime }(r),  \notag \\
v(r,t) &=&\frac{2}{\left( t_{0}\pm t\right) }\frac{m(r)}{m^{\prime }(r)}%
,\Omega (t)=\pm \frac{\beta _{0}\omega _{0}}{t_{0}\pm t}.
\end{eqnarray}

We rescale the radial coordinate $r$, the mass $m(r)$, the density $\rho
(r,t)$ and the velocity $v(r,t)$ according to the transformations
\begin{equation*}
\hspace{-0.3cm}r=\frac{R_{s}}{\pi }\eta ,\;\;m(r)=M_{S}m_{0}(\eta ),
\end{equation*}%
where $R_{S}=\sqrt{\pi u_{0}/2G}$ and $M_{S}=\left( 4/\pi \right)
R_{S}^{3}\;\rho _{c}$ are the radius and the mass of the static condensate,
while $\eta $ and $m_{0}(\eta )$ are the dimensionless radial coordinate and
the dimensionless mass function, respectively. We denote $\gamma
_{0}^{2}=\omega _{0}^{2}/6\pi ^{2}G\rho _{c}$. Moreover, we also fix in the
following the constant $\beta _{0}^{2}$ as
\begin{equation}
\beta _{0}^{2}=\frac{1}{4\pi ^{2}G\rho _{c}}.
\end{equation}

Hence for the physical parameters of the collapsing/expanding Bose-Einstein
Condensate dark matter halo we obtain the expressions
\begin{eqnarray}
\hspace{-0.1cm}M(\eta ,t) &=&\frac{M_{S}}{4\pi ^{2}G\rho _{c}}\frac{%
m_{0}(\eta )}{\left( t_{0}\pm t\right) ^{2}},\rho (\eta ,t)=\frac{1}{4\pi G}%
\frac{1}{\left( t\pm t_{0}\right) ^{2}}\theta \left( \eta \right) ,  \notag
\\
\hspace{-0.1cm}v(\eta ,t) &=&\frac{2}{\left( t\pm t_{0}\right) }\frac{R_{S}}{%
\pi }V(\eta ),\Omega (t)=\pm \frac{\omega _{0}}{\sqrt{4\pi ^{2}G\rho _{c}}}%
\frac{1}{t_{0}\pm t}.  \notag \\
&&
\end{eqnarray}

The dimensionless spatial density distribution of the density $\theta \left(
\eta \right) $ and velocity $V(\eta )$ of the Bose-Einstein Condensate dark
matter halo can be obtained as
\begin{equation}
\theta (\eta )=\frac{1}{\eta ^{2}}m_{0}^{\prime }(\eta ),V(\eta )=\frac{%
m_{0}(\eta )}{m^{\prime }(\eta )}.
\end{equation}

Then, in the new dimensionless variables, Eq.~(\ref{95}), describing the
mass profile of the evolving dark matter halo, takes the form
\begin{eqnarray}  \label{98}
&&m_0^{\prime \prime }(\eta)-\frac{2}{\eta}m_0^{\prime
}(\eta)+m_0(\eta)-\gamma _0^2\eta ^3=  \notag \\
&&2\eta ^2\Bigg\{ \frac{m_0(\eta)}{m_0^{\prime }(\eta)}-\frac{d}{d\eta}\left[%
\frac{m_0(\eta)}{m_0^{\prime }(\eta)}\right]^2 \Bigg\},
\end{eqnarray}
or, equivalently,
\begin{eqnarray}  \label{99}
&&\left[1-4\eta ^2\frac{m_0^2(\eta)}{m_0^{\prime \;3}(\eta)}\right]%
m_0^{\prime \prime }(\eta)-\frac{2}{\eta}m_0^{\prime }(\eta)+  \notag \\
&&\left[1+\frac{2\eta ^2}{m_0^{\prime }(\eta)}\right]m_0(\eta)-\gamma
_0^2\eta ^3=0.
\end{eqnarray}

For the central density of the condensate at the beginning of the contracting/expanding phase we obtain $\rho (0,0)=\rho _c=\theta (0)/4\pi Gt_0^2$,
which gives $\theta (0)=\theta _0=4\pi G\rho _ct_0^2$.

\subsection{The first order approximation}

In the zeroth order approximation, we can neglect the right hand side in
Eq.~(\ref{98}), which thus takes the form
\begin{equation}
m_0^{\prime \prime }(\eta)-\frac{2}{\eta}m_0^{\prime
}(\eta)+m_0(\eta)-\gamma _0^2\eta ^3= 0,
\end{equation}
and has the general solution
\begin{eqnarray}
m_0^{(0)}(\eta)&=&\gamma _0^2 \eta ^3-\sqrt{\frac{2}{\pi }} \Bigg[\left(C_2
\eta-C_1\right) \sin (\eta)+  \notag \\
&& \left(C_1 \eta+C_2\right) \cos (\eta)\Bigg],
\end{eqnarray}
where $C_1$ and $C_2$ are arbitrary constants of integration. The boundary
condition $m_0^{(0)}(0)=0$ gives $C_2=0$. For the dimensionless density we
obtain
\begin{equation}
\theta ^{(0)}(\eta)=\sqrt{\frac{2}{\pi }} C_1\frac{ \sin (\eta)}{\eta}+3
\gamma _0^2.
\end{equation}
The boundary condition $\theta ^{(0)}(0)=\theta _0$ gives $C_1=\sqrt{\pi/2}\left(\theta _0-3\gamma
_0^2\right)$. Hence we obtain the zeroth order solution of Eq.~(%
\ref{98}) as
\begin{equation}
m_0^{(0)}(\eta)=\gamma _0^2 \eta ^3+\left(\theta _0-3 \gamma _0^2\right)\left[ \sin
\eta- \eta \cos \eta \right].
\end{equation}
The density is given by
\begin{equation}
\theta ^{(0)}(\eta)=\frac{\left(\theta _0-3 \gamma _0^2\right) \sin \eta}{\eta}+3
\gamma _0^2.
\end{equation}

To obtain the solution of Eq.~(\ref{98}) in the first order of approximation
we substitute in its right hand side $m(\eta)$ by $m_0^{(0)}(\eta)$, thus
obtaining
\begin{eqnarray}
&&h(\eta)=2\eta ^2\Bigg\{ \frac{m_0(\eta)}{m_0^{\prime }(\eta)}-\frac{d}{%
d\eta}\left[\frac{m_0(\eta)}{m_0^{\prime }(\eta)}\right]^2 \Bigg\}=\frac{2
\eta ^3}{9}+  \notag \\
&&\left(\frac{2 \gamma _0^2}{9}-\frac{2}{27}\right) \eta ^5+\frac{%
\left(-1197 \gamma _0^4+663 \gamma _0^2-88\right) \eta ^7}{4725}+O\left(\eta
^8\right).  \notag \\
\end{eqnarray}

Hence in this approximation Eq.~(\ref{98}) becomes
\begin{eqnarray}
\hspace{-0.5cm} &&m_{0}^{\prime \prime }(\eta )-\frac{2}{\eta }m_{0}^{\prime
}(\eta )+m_{0}(\eta )=\left( \gamma _{0}^{2}+\frac{2}{9}\right) \eta ^{3}+
\notag \\
\hspace{-0.5cm} &&\left( \frac{2\gamma _{0}^{2}}{9}-\frac{2}{27}\right) \eta
^{5}+\frac{\left( -1197\gamma _{0}^{4}+663\gamma _{0}^{2}-88\right) \eta ^{7}%
}{4725},
\end{eqnarray}%
with the general solution given by
\begin{eqnarray}
\hspace{-0.8cm}m_{0}^{(1)}(\eta ) &=&-\frac{7\left( 9576\gamma
_{0}^{4}-5139\gamma _{0}^{2}+574\right) \eta ^{3}}{945}+  \notag \\
\hspace{-0.8cm}&&\frac{14\left( 3\gamma _{0}^{2}-1\right) \left( 798\gamma
_{0}^{2}-151\right) \eta ^{5}}{4725}+  \notag \\
\hspace{-0.8cm}&&\frac{\left( -1197\gamma _{0}^{4}+663\gamma
_{0}^{2}-88\right) \eta ^{7}}{4725}-  \notag \\
\hspace{-0.8cm}&&\sqrt{\frac{2}{\pi }}\left[ \left( c_{4}\eta -c_{3}\right)
\sin \eta +\left( c_{3}\eta +c_{4}\right) \cos \eta \right] ,
\end{eqnarray}%
where $c_{3}$ and $c_{4}$ are arbitrary constants of integration. The
boundary condition $m_{0}^{(1)}(0 )=0$ gives $C_4=0$, while the density
profile of the Bose-Einstein Condensate dark matter halo is obtained as
\begin{eqnarray}
\hspace{-0.2cm}\theta ^{(1)}(\eta)&=&\sqrt{\frac{2}{\pi }} c_3\frac{ \sin
\eta}{\eta}+\frac{\left(-1197 \gamma _0^4+663 \gamma _0^2-88\right) \eta ^4}{%
675} +  \notag \\
\hspace{-0.2cm} && \frac{2}{135} \left(2394 \gamma _0^4-1251 \gamma
_0^2+151\right) \eta ^2+  \notag \\
\hspace{-0.2cm} && \frac{1}{45} \left(-9576 \gamma _0^4+5139 \gamma
_0^2-574\right).
\end{eqnarray}
By imposing the boundary condition $\theta ^{(1)}(0)=1$ we obtain $C_3=
\sqrt{\pi /2} \left(9576 \gamma _0^4-5139 \gamma _0^2+619\right)/45$. Hence
for the spatial mass and density profiles of the dynamically evolving
Bose-Einstein Condensate dark matter halos we obtain, in the first order
approximation, the expressions
\begin{eqnarray}
m_{0}^{(1)}(\eta )&=&\frac{\left(-1197 \gamma _0^4+663 \gamma _0^2-88\right)
\eta ^7}{4725}+  \notag \\
&& \frac{2}{675} \left(2394 \gamma _0^4-1251 \gamma _0^2+151\right) \eta ^5+
\notag \\
&& \frac{1}{135} \left(-9576 \gamma _0^4+5139 \gamma _0^2-574\right) \eta ^3+
\notag \\
&& \frac{1}{45} \left(9576 \gamma _0^4-5139 \gamma _0^2+619\right) \sin \eta
-  \notag \\
&& \frac{1}{45} \left(9576 \gamma _0^4-5139 \gamma _0^2+619\right) \eta \cos
\eta,
\end{eqnarray}
and
\begin{eqnarray}
\theta ^{(1)}(\eta)&=& \frac{1}{675} \left(-1197 \gamma _0^4+663 \gamma
_0^2-88\right) \eta ^4+  \notag \\
&& \frac{2}{135} \left(2394 \gamma _0^4-1251 \gamma _0^2+151\right) \eta ^2+
\notag \\
&& \frac{\left(9576 \gamma _0^4-5139 \gamma _0^2+619\right) \sin \eta}{45
\eta}+  \notag \\
&& \frac{1}{45} \left(-9576 \gamma _0^4+5139 \gamma _0^2-574\right),
\end{eqnarray}
respectively. In the first order approximation the spatial velocity profile
of the evolving Bose-Einstein Condensate dark matter halo can be obtained as
$V^{(1)}(\eta)=m_{0}^{(1)}(\eta )/dm_{0}^{(1)}(\eta )/d\eta $.

The above results simplify considerably in the case of the nonrotating
condensate dark matter halos, with $\gamma _0^2=0$. Hence, for a radially
expanding/contracting Bose-Einstein Condensate confined by its own
gravitational field we obtain
\begin{equation}
m_{0}^{(1)}(\eta )=-\frac{88 \eta ^7}{4725}+\frac{302 \eta ^5}{675}-\frac{%
574 \eta ^3}{135}+\frac{619 }{45}\left(\sin \eta- \eta \cos \eta \right),
\end{equation}
\begin{equation}
\theta ^{(1)}(\eta)=\frac{619 \sin \eta }{45 \eta}-\frac{2}{675} \left(44
\eta ^4-755 \eta ^2+4305\right),
\end{equation}
\begin{eqnarray}
\hspace{-1.3cm}&&V^{(1)}(\eta)=\frac{1}{\eta}+  \notag \\
\hspace{-1.3cm}&&\frac{88 \eta ^6-2730 \eta ^4+30660 \eta ^2+64995 \cos
\eta-60270}{7 \left(88 \eta ^5-1510 \eta ^3+8610 \eta -9285 \sin \eta \right)%
}.
\end{eqnarray}

In this approximation the mass and density are finite at the origin $\eta =0$%
, but the velocity diverges at the center of the contracting condensate dark
matter halo.

\subsection{Exact numerical profiles}

Eq.~(\ref{99}) can be reformulated mathematically as a first order dynamical
system given by
\begin{equation}  \label{102}
\frac{dm_0}{d\eta}=u,
\end{equation}
\begin{eqnarray}  \label{103}
\frac{du}{d\eta}&=&\frac{2}{\eta}\frac{u}{1-4\eta ^2m_0^2/u^3}-\frac{%
\left(1+2\eta ^2/u\right)m_0}{1-4\eta ^2m_0^2/u^3}+  \notag \\
&&\frac{\gamma _0^2\eta ^3}{1-4\eta ^2m_0^2/u^3}.
\end{eqnarray}

The system of equations (\ref{102}) and (\ref{103}) must be integrated with
the initial conditions $m_0(0)=0$ and $u(0)=0$, respectively. As for the
dimensionless density $\theta$ and velocity $V(\eta)$ of the Bose-Einstein
Condensate dark matter they can be represented as
\begin{equation}
\theta (\eta)=\frac{1}{\eta ^2}u, V(\eta)=\frac{m_0}{u}.
\end{equation}

\begin{figure}[!htb]
\centering
\includegraphics[scale=0.65]{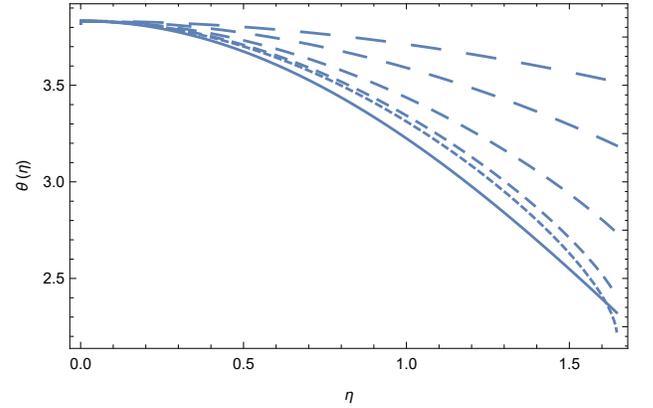}
\caption{Variation of the dimensionless density profile  $\theta (\eta)$ as a function of the dimensional radial coordinate $\eta$  of a collapsing rotating Bose-Einstein Condensate dark matter halo for different values of the parameter $\gamma _0$: static case with $\gamma _0=0$, (solid curve), and dynamical evolution with $\gamma _0=0$ (dotted curve), $\gamma _0=0.25$ (short dashed curve), $\gamma _0=0.50$ (dashed curve), $\gamma _0=0.75$ (long dashed curve), and $\gamma _0=0.90$ (long dashed curve), respectively. To numerically integrate the  system of equations (\ref{102}) and (\ref{103}) we have used the initial conditions $m_0\left(\eta _0\right)=\left(4/\pi \right) \eta _0^3$ and $u\left(\eta _0\right)=m_0'\left(\eta _0\right)=\left(12/\pi \right) \eta _0^2$, with $\eta _0=10^{-3}$.}
\label{fig1}
\end{figure}

\begin{figure}[!htb]
\centering
\includegraphics[scale=0.65]{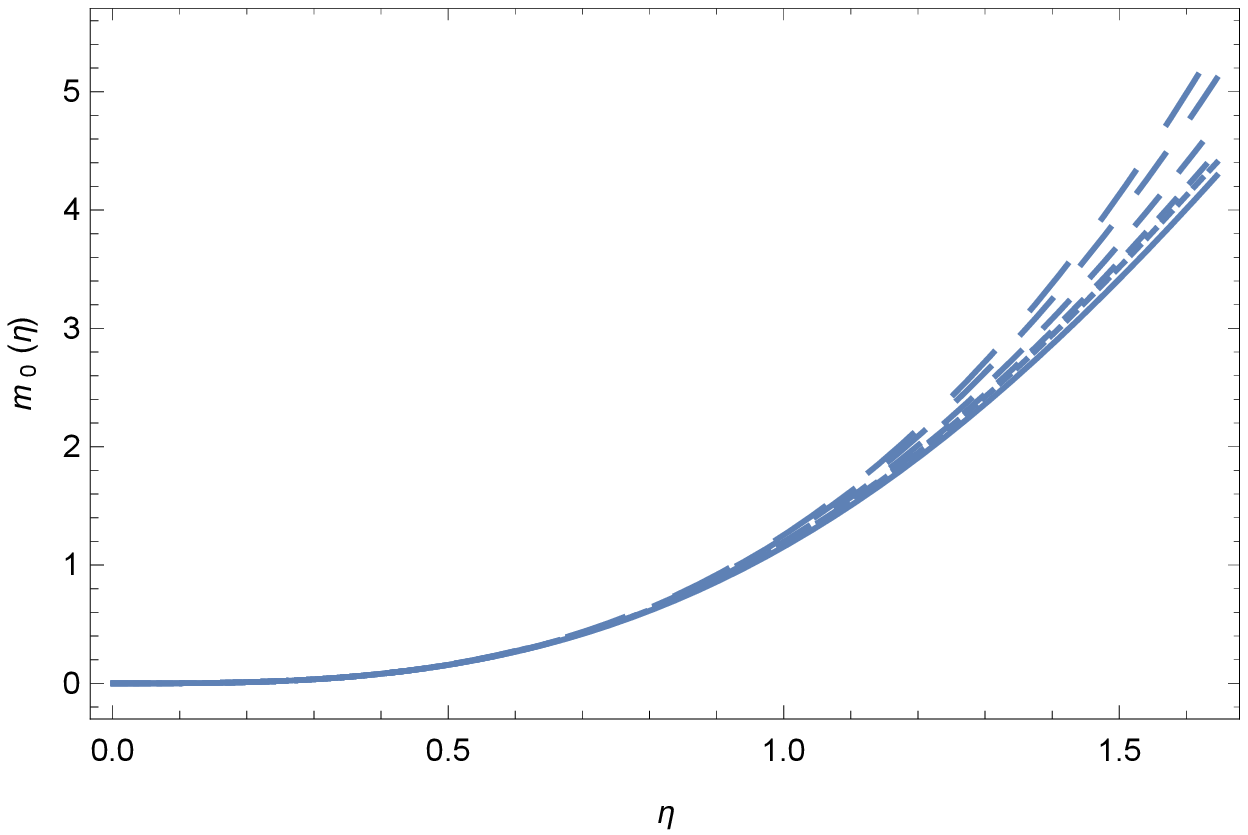}
\caption{Variation of the dimensionless mass profile  $m_0(\eta)$ as a function of the dimensional radial coordinate $\eta$  of a collapsing rotating Bose-Einstein Condensate dark matter halo for different values of the parameter $\gamma _0$: static case with $\gamma _0=0$, (solid curve), and dynamical evolution with $\gamma _0=0$ (dotted curve), $\gamma _0=0.25$ (short dashed curve), $\gamma _0=0.50$ (dashed curve), $\gamma _0=0.75$ (long dashed curve), and $\gamma _0=0.90$ (long dashed curve), respectively. To numerically integrate the  system of equations (\ref{102}) and (\ref{103}) we have used the initial conditions $m_0\left(\eta _0\right)=\left(4/\pi \right) \eta _0^3$ and $u\left(\eta _0\right)=m_0'\left(\eta _0\right)=\left(12/\pi \right) \eta _0^2$, with $\eta _0=10^{-3}$.}
\label{fig2}
\end{figure}

\begin{figure}[!htb]
\centering
\includegraphics[scale=0.65]{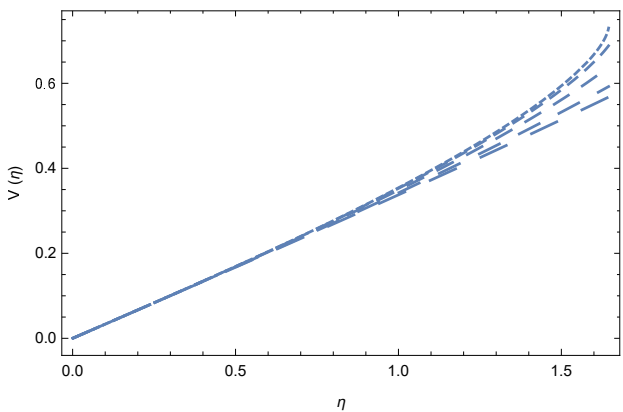}
\caption{Variation of the dimensionless velocity profile  $V(\eta)$ as a function of the dimensional radial coordinate $\eta$  of a collapsing rotating Bose-Einstein Condensate dark matter halo for different values of the parameter $\gamma _0$ during its dynamical evolution for $\gamma _0=0$ (solid curve), $\gamma _0=0.25$ (dotted curve), $\gamma _0=0.50$ (short dashed curve), $\gamma _0=0.75$ (dashed curve), and $\gamma _0=0.90$ (long dashed curve), respectively. To numerically integrate the  system of equations (\ref{102}) and (\ref{103}) we have used the initial conditions $m_0\left(\eta _0\right)=\left(4/\pi \right) \eta _0^3$ and $u\left(\eta _0\right)=m_0'\left(\eta _0\right)=\left(12/\pi \right) \eta _0^2$, with $\eta _0=10^{-3}$.}
\label{fig3}
\end{figure}

The variations of the dimensionless spatial density, mass and velocity profiles of the contracting/expanding Bose-Einstein Condensate dark matter halos are represented in Figs.~\ref{fig1}-\ref{fig3}, respectively. To numerically integrate the system of equations  (\ref{102}) and (\ref{103}) we have used the initial conditions $m_0\left(\eta _0\right)=\left(4/\pi \right) \eta _0^3$ and $m_0'\left(\eta _0\right)=\left(12/\pi \right) \eta _0^2$, with $\eta _0=10^{-3}$. The spatial density profile of the dark matter halo during its dynamical evolution is represented in Fig.~\ref{fig1}. The static solution in the absence of rotation is also represented, as the lower curve in the figure. In all cases the density profiles are described by monotonically decreasing functions of the dimensionless radius $\eta$. The effects of the rotation of the halo can be clearly seen, and they lead to a significant impact on the density profiles. Near the origin, for $\eta $ in the range $0<\eta <0.25$, the density profiles are basically indistinguishable, but for larger values of $\eta$ the effects of the rotation modify the overall density distribution. The impact of the rotation on the spatial mass distribution of the collapsing/expanding dark matter halos is presented in Fig.~\ref{fig2}. The mass profiles are characterized by linearly increasing functions of $\eta$. For $\eta $ in the range $0<\eta <1.1$, the mass distribution of the rotating collapsing/expanding condensate dark matter halo essentially coincides with the static mass distribution. The effects of the rotation become important for large values of the radius, once we are approaching the lower density regions at the outer boundary of the halo. The rotation can lead to a significant increase in the total mass of the halo, the increase being of the order of 20\% for $\gamma _0=0.9$.

The spatial velocity profiles of the dynamically evolving condensate dark matter halos are presented in Fig.~\ref{fig3}. Similarly to the mass profile, the spatial distribution of the velocity is a monotonically increasing function of $\eta$. For $\eta $ in the range $0<\eta <1.1$, the velocity profiles are (almost) identical, and the effects of rotation are extremely small. For small rotation velocities the velocity can be well approximated by a linear function of $\eta $, so that $V(\eta)\propto \eta$. However, for larger values of the dimensionless radial coordinate, the effects of the rotation become important at large distances from the halo center, and the velocity - radial coordinate relations is not linear anymore.

\section{Discussions and final remarks}\label{sect4}

In the present paper we have investigated some of the possible physical and
astrophysical consequences of the instabilities in a Bose-Einstein
Condensate dark matter halo. We have considered two distinct, but
interrelated topics, the Jeans stability of condensate dark matter clouds,
and the dynamics of the gravitational collapse that follows once the real
size of the gravitationally confined system exceeds the Jeans scales.

The Bose-Einstein Condensates are generally described by the
Gross-Pitaevskii equation, which gives the ground state of a quantum system
of identical bosons, and which is obtained by using the Hartree-Fock approximation and the pseudopotential interaction model. In the
present approach we assume that dark matter is a Bose-Einstein
Condensate of a gas of bosons, which are in the same quantum state, and thus
can be described by the same wavefunction. Hence, we interpret the galactic
dark matter halos as huge quantum systems, whose properties can be described
by a single wave function. To simplify our formalism we also adopt the
assumption that dark matter is at zero temperature. Of course the presence
of the excitations and of nonzero temperature effects may have important
effects on the properties of condensate dark matter.

An important mathematical and physical property of the Gross-Pitaevskii
equation is that, similarly to the standard Schr\"{o}dinger equation, it
admits a hydrodynamical representation, easily obtainable after representing
the wave function in the Madelung variables. It turns out that in the
hydrodynamic representation the Bose-Einstein Condensates can be described
as a perfect fluid, and its properties are characterized by the local
density and local velocity only. The fluid satisfies a continuity and an
Euler type equations, which contains the contributions of the quantum
pressure $p=u_0\rho ^2$, generated by the self-interaction of the field, and
of the quantum potential $V_Q$, both of these terms being essentially
quantum mechanical in their origin. The velocity $\vec{v}$ of the quantum
fluid is related to the phase $S$ of the wave function by the relation $\vec{%
v}=\left(\hbar /m\right)\nabla S$, and the flow of the condensate fluid is
rotationless, as long as $S$ contains no singularities, as, for example, in
a vortex. In the present approach we have also allowed the presence of
vortices in the dark matter halo. The simplest method to create vortices in
a condensate is through their rotation. Hence, under the natural assumption
that Bose-Einstein Condensate dark matter halos are rotating, one must also
take into account the existence of vortices in the system. A Bose-Einstein
Condensate dark matter halo can rotate only due to the existence of
quantized vortex lines. When the rotation frequency $\Omega $ of the halo
exceeds a critical value $\Omega _c$, vortex nucleation occurs. In the
present paper we have adopted the simple relation (\ref{20}) between the
angular velocity of the rotating condensate dark matter halos, and their
vortex number density, which indicates that the quicker the dark matter halo
rotates, the higher the number of vortices. On the other hand in rotating
gravitationally trapped Bose-Einstein Condensate dark matter halos the
nucleation of vortices may also be a result of the instabilities of
collective excitations.

In the present manuscript we have used the term turbulence in the sense that it is common in the physics of quantum fluids, meaning the presence in the fluid of quantized vortices \cite{turb1,turb2}. In this nomenclature turbulence is more related to
the quantization of vortices than to the lack of viscosity of superfluids. On the other hand  the term turbulence may also be interpreted as describing a phase of temporally and spatially disordered fluid motion,
characterized by a large number of degrees of freedom interacting essentially  nonlinearly \cite{turb2}. The nonlinear
interaction is usually a consequence of the presence of the term $\left(\vec{v}\cdot \nabla\right)\vec{v}$  in the Euler equation.

The hydrodynamical description of the rotating condensate dark matter halos,
confined by their own gravitational fields, opens the possibility of the
study of the gravitational perturbations in galactic and extragalactic
systems. By assuming some standard initial conditions, the linearization of
the hydrodynamic equations describing the flow of the dark matter clouds
leads to the dispersion relation (\ref{32}), which is exact in the sense
that it also includes both the effects of the rotation and of the quantum
force. In the limit when the rotation and the quantum force effects are
neglected, Eq.~(\ref{32}) reduces to the standard form of the Jeans
frequency,
\begin{equation}
\omega ^2=v_s^2\vec{k}^2-4\pi G\rho _0,
\end{equation}
and to the Jeans radius $R_J=\left(\sqrt{\pi/G\rho _0}\right)v_s$. The Jeans
radius essentially depends on the speed of sound in the given astrophysical
medium, and it is
a measure of the linear dimensions of the condensations which will form in the
medium on account of the gravitational instability. For matter obeying the ideal gas equation of state $p=\left(k_B
T/m\mu \right)$, where $T$ is the temperature of the gas, $\mu $ is the mean
molecular weight, and $k_B$ is the Boltzmann constant, we have $v_s=\sqrt{%
k_BT/m\mu}$, and $R_J=\sqrt{\pi k_BT/G\rho _0m\mu}$. For an ideal gas the
Jeans length is proportional to the temperature of the system, and inversely
proportional to the density. On the other hand for a Bose-Einstein
Condensate in the same approximation we obtain $R_J=\sqrt{4\pi ^2\hbar
^2a/Gm^3}$, and expression that depends only on the fundamental physical
characteristics of the dark matter halo (particle mass and scattering
length), and on the Planck constant. Hence, for a Bose-Einstein Condensate
at zero temperature the Jeans radius has an universal expression,
independent on the macroscopic properties of the medium.

In the opposite limit of zero speed of sound (or negligible quantum
pressure), the dispersion relation takes the form
\begin{equation}
\omega ^{2}=\frac{\hbar ^{2}}{4m^{2}}\vec{k}^{4}-4\pi G\rho _{0}\left( 1+%
\frac{\vec{\Omega}_{0}^{2}}{\pi G\rho _{0}}\right) ,
\end{equation}%
giving a Jeans wave number given by
\begin{equation}
\left\vert \vec{k}\right\vert _{J}=\left[ \frac{16\pi G\rho _{0}m^{2}\left(
1+\vec{\Omega}_{0}^{2}/\pi G\rho _{0}\right) }{\hbar ^{2}}\right] ^{1/4},
\label{136}
\end{equation}%
and a Jeans radius of the order of
\begin{equation}
R_{J}=\left[ \frac{\pi ^{3}\hbar ^{2}}{G\rho _{0}m^{2}\left( 1+\vec{\Omega}%
_{0}^{2}/\pi G\rho _{0}\right) }\right] ^{1/4}.  \label{137}
\end{equation}

In the absence of rotation, Eqs. (\ref{136}) and (\ref{137}) give the Jeans
wave number and the Jeans radius corresponding to the Schr\"{o}%
dinger-Poisson model, described by the system of equations \cite{SP1, SP2,SP3,SP4,SP5}
\begin{equation}
i\hbar \frac{\partial \Psi \left( \vec{r},t\right) }{\partial t}=-\frac{%
\hbar ^{2}}{2m}\Delta \Psi \left( \vec{r},t\right) +mV_{g}\left( \vec{r}%
,t\right) \Psi \left( \vec{r},t\right) ,
\end{equation}
\begin{equation}
\Delta V_{g}\left( \vec{r},t\right) =4\pi Gm^{2}\left\vert \Psi \left( \vec{r%
},t\right) \right\vert ^{2},
\end{equation}%
where $V_{g}\left( \vec{r},t\right) $ is the gravitational potential. The
essential difference between the Schr\"{o}dinger-Poisson and
Gross-Pitaevskii-Poisson models of dark matter is that in the former the
effect of the particle self-interaction is neglected, which implies that in
the Schr\"{o}dinger-Poisson model there is no quantum pressure term $p$, and the
speed of sound in the dark matter halo is zero. By assuming a zero
rotational velocity, for the Jeans radius and Jeans mass of the dark matter
halo in the Schr\"{o}dinger-Poisson model we obtain the relations
\begin{eqnarray}
R_{J}&=&\left[ \frac{\pi ^{3}\hbar ^{2}}{G\rho _{0}m^{2}}\right]
^{1/4}=3.57\times 10^{12}\times \left( \frac{m}{\mathrm{meV}}\right)
^{-1/2}\times  \notag \\
&&\left( \frac{\rho _{0}}{10^{-24}\;\mathrm{g/cm^{3}}}\right) ^{-1/4}\;%
\mathrm{cm},
\end{eqnarray}
and
\begin{eqnarray}
\hspace{-0.5cm}&&M_{J}=\left(\frac{2^{8/3}\pi ^{10/3}\hbar ^2}{3^{4/3}Gm^2}%
\right)^{3/4}\rho _0^{1/4}=  \notag \\
\hspace{-0.5cm}&&9.532\times 10^{-20}\times \left( \frac{m}{\mathrm{meV}}%
\right) ^{-3/2}\left( \frac{\rho _{0}}{10^{-24}\;\mathrm{g/cm^{3}}}%
\right)^{1/4} M_{\odot },  \notag \\
\end{eqnarray}
respectively. In the Schr\"{o}dinger-Poisson model all the physical and
astrophysical properties of the dark matter halos are fixed by a single
parameter, the mass of the dark matter particle. In order to obtain
realistic astrophysical results applicable to the galactic or extra-galactic
scales very small particle masses are required, a problem the Bose-Einstein
Condensate dark matter model does not have.

We have also discussed the effects of the rotation on the condensate dark
matter halos. In realistic astrophysical situations the condition $\vec{%
\Omega}^{2}/\pi G\rho _{0}>>1$ may be satisfied. In this case the Jeans wave
number is proportional to the angular velocity of the halo, while the Jeans
radius and Jeans mass are inversely proportional to $\left\vert \vec{\Omega}%
_{0}\right\vert $, and $\left\vert \vec{\Omega}_{0}\right\vert ^{3}$,
respectively. For large angular velocities the Jeans radius is very small,
of the order of $10^{-2}$ kpc, while the corresponding Jeans mass is of the
order of $90M_{\odot}$. These radius and mass values may correspond to
stellar mass black holes, or other types of small (in an astrophysical
sense) compact objects. In the case of ordinary baryonic matter the condition  of instability
for waves propagating perpendicularly to the axis of rotation is $k_J^2v_s^2<4\pi G\rho _0-4\Omega ^2$ \cite{Chand2}.

One constraint on the ratio and scattering length of the dark matter
particle can be obtained from the condition $2GM_{J}/c^{2}R_{J}<1$,
requiring that dark matter halos cannot become black holes. This condition
is essentially general relativistic, but we may assume that it is also valid
in Newtonian mechanics. By adopting for the Jeans radius and mass the
expressions (\ref{Jeans1}) and (\ref{34}), respectively, we obtain the
constraint
\begin{equation}
\frac{a}{m^3}<\frac{3c^2}{32\pi ^3\hbar ^2\rho _0}=2.449\times 10^{96}
\left( \frac{\rho _{0}}{10^{-24}\;\mathrm{g/cm^{3}}}\right) ^{-1}\;\mathrm{%
cm/g^3}.
\end{equation}

Once the numerical values of the critical parameters of the condensate dark matter halos are reached, an instability develops in the system. One can interpret, from a simple physical point of view, the collapse of
the Bose-Einstein Condensate dark matter halos as follows. Once the number of particles $N$ in the galactic halo becomes sufficiently
high, so that $N > N_c$, where $N_c$ is a critical particle number, the
attractive inter-particle energy exceeds the quantum
pressure, and the dark matter halo begins to collapse. During the collapsing stage, the density of dark matter particles increases in
the  vicinity of the galactic center.

To study the dynamical evolution of the halo one need to solve the full system of hydrodynamic equations describing the time and space evolution of the Bose-Einstein Condensate. In the Thomas-Fermi approximation the basic equations describing the dynamics of the condensate coincide with the continuity and Euler equations of classical hydrodynamics. The effect of the rotation was also included by means of a term, averaged over the angle $\theta $, proportional to the radial coordinate, and with a time-dependent angular frequency. In order to preserve spherical symmetry we assume that the rotation of the condensate is slow. The effect of the gravitational field is taken into account via the Poisson equation, which relates the gravitational potential of the confined condensate to its density, which is essentially quantum in its nature. After integrating the Poisson equation, the evolution equations can be reduced to a system of two coupled nonlinear partial differential equations, which can subsequently be reduced to a single nonlinear second order partial differential equations for the mass of the condensate $M(r,t)$. For many hydrodynamic flow models in gravitational fields, with matter obeying a polytropic equation of state the equations of motion admit semi-analytical self-similar or homologously collapsing solutions \cite{hom1, hom2,hom3,hom4,hom5}.

 When these solutions are valid scale radial profiles of thermodynamic variables remain time invariant.  The homologous solutions are stable against
linear nonspherical perturbations, a result which can explains the remarkable stability of the galactic dark matter halos. However, in the case of index $n=1$ polytropes no self-similar solution does exist, and the solutions of the equations of motion cannot be written in the form $\Phi (r,t)=f(t)g\left(r/t\right)$ \cite{hom1,hom2,hom3,hom4,hom5}. This means that during their evolution the physical quantities describing Bose-Einstein Condensate dark matter halos {\it are not scale and time invariant}. However an exact semi-analytical solution of the evolution equations of the Bose-Einstein Condensates self-confined by the gravitational field can be obtained by assuming the separability of the mass variable in the for $M(r,t)=f(t)m(r)$. The function $f(t)$ has the analytical form $f(t)=\left(t_0\pm t\right)^{-2}/4\pi ^2 G\rho _c$, where the minus sign corresponds to the collapse of the condensate, while the plus sign describes the expansion of the condensate dark matter halo. As for the mass function $m(r)$, time evolution generates a universal condensate dark matter profile, which in the absence of rotation just slightly modifies the static mass and density profiles. However, once the rotation is taken into account, the mass and density profiles are significantly modified as functions of the angular velocity.

The obtained solution describes both an expanding and a collapsing phase. The expanding phase corresponds to the choice of plus sign in the function $f(t)$, so that $f(t)=\left(t_0+ t\right)^{-2}/4\pi ^2 G\rho _c$. The Bose-Einstein Condensate disperses in the space, with the local values of the physical parameters decreasing in time. In the large time limit the density as well as the radial and angular velocities tend to zero, and the Bose-Einstein Condensate reaches a vacuum state. The minus sign in the function $f(t)$ describes the collapse of the condensate. Both solutions are singular, for the expanding case the singularity occurs at the initial moment $t=0$ (by a rescaling of the time variable one can remove the constant $t_0$ from the results, $t\rightarrow t+t_0$), while in the case of the gravitational collapse the singularity is reached at the finite time $t_0$. In order to obtain a rough estimate of $t_0$, we assume that the mass distribution $m(r)$ during the collapsing phase does not differ much from the stationary one. By series expanding Eq.~(\ref{93}) we obtain $m(r)\approx M_0(r)\approx \left(4\pi /3\right)\rho _c r^3$, an expression which allows us to estimate the velocity of the collapsing Bose=Einstein Condensate dark matter halos as
\be\label{147}
v(r,t)\approx \frac{2}{t_0-t}\frac{r}{3}.
\ee

At the beginning of the gravitational collapse at $t=0$, the extension of the condensate cloud is given by its Jeans radius $R_J$, which we approximate by Eq.~(\ref{Jeans1}). Moreover, we assume that the initial velocity of the collapsing boundary can be written as $v\left(R_J,0\right)=\alpha c$, where $\alpha$ is a constant. Then for the collapse time $t_9$ of the condensate halo we obtain
\be
t_0\approx \frac{2R_J}{3\alpha c}=\frac{23.30}{\alpha }\times 10^{11}\times \left(\frac{m}{\mathrm{meV}}\right)^{-3/2}\left(\frac{a}{%
10^{-3}\;\mathrm{fm}}\right)^{1/2}\;\mathrm{s}.
\ee

An alternative estimate of the collapse time can be obtained by assuming that at the beginning of the collapse the total mass of the condensate dark matter halo $M\left(R_J,0\right)=m\left(R_J\right)$ is given by the Jeans mass (\ref{34}). Then we obtain for the collapse time the simple expression $t_0\approx 1/2\pi \sqrt{G\rho _c}$.

It is interesting to note that,  at least in the first order of approximation, the velocity of the condensate dark matter halo has the mathematical form of the cosmological Hubble law, $v(t)=H(t)r$, with the Hubble function given by $H(t)=2/\left(t_0\pm t\right)$.  Moreover, as shown by the numerical simulations presented in Fig.~\ref{fig3}, for small angular velocities the simple linear proportionality between the dimensionless velocity $V$ and the radial dimensionless coordinate $\eta $ is valid during the entire phase of the collapsing/expanding evolution.

A large number of cosmological and astrophysical observations, including the behavior of the  galactic rotation curves, or the virial mass discrepancy in galaxy clusters points  towards the possibility of the existence of dark matter in the Universe. If dark matter is present in the form of bosonic particles, then the possibility that it is in the form of a Bose-Einstein Condensate is strongly supported by the laws of quantum physics that show that a phase transition to a condensate state must occur, once the temperature reaches its critical value. A major adjustment in our interpretation of the basic principles of astrophysics and cosmology will be required if further observations  would confirm this hypothesis. In the present papers we have introduced some basic theoretical tools that could help in the analysis and interpretation of the large scale structure formation in the presence of Bose-Einstein Condensate dark matter.

\appendix

\section{No self-similar contraction/expansion of Bose-Einstein Condensate dark matter
halos}\label{appa}

Eqs.~(\ref{s4}) and (\ref{s5}) are invariant under the time reversal
operations $t\rightarrow -t$, $v\rightarrow -v$, and $\rho \rightarrow \rho$%
, $M\rightarrow M$, respectively. Hence, without any loss of generality, one
can restrict the analysis to the range $0 \leq t < \infty$ of the time
variable. In order to investigate the possibility of the existence of a
self-similar solution that describes the collapse of Bose-Einstein
Condensate dark matter halos, we introduce first a similarity variable $\xi $%
, and assume that each physical quantity scales according to the following
rules,
\begin{eqnarray}  \label{A1}
r&=&\alpha (t)\xi, v(r,t)=\beta (t)u\left(\xi \right),  \notag \\
\rho (r,t)&=&\gamma (t) \theta \left(\xi \right), M(r,t)=\eta (t )m\left(\xi
\right),
\end{eqnarray}
where $\alpha (t)$, $\beta (t)$, $\gamma (t)$ and $\eta (t)$ are continuous
functions of the time variable $t$ only.

Then we obtain first
\begin{equation}
\frac{dm(\xi)}{d\xi}=4\pi \frac{\gamma (t)\alpha ^3(t)}{\eta (t)}\xi
^2\theta (\xi).
\end{equation}

In the self-similar variables the continuity equation (\ref{s4}) gives
\begin{eqnarray}
m(\xi)&=&\frac{\eta (t)}{\dot{\eta}(t)\alpha (t)}\left[\dot{\alpha}%
(t)\xi-\beta (t)u(\xi)\right]\frac{dm(\xi)}{d\xi}=  \notag \\
&&4\pi\frac{\gamma (t)}{\dot{\eta}(t)}\alpha ^2(t)\xi ^2\theta (\xi)\left[%
\dot{\alpha}(t)\xi-\beta (t)u(\xi)\right],
\end{eqnarray}
while the Euler equation (\ref{s5} takes the form
\begin{eqnarray}
\hspace{-0.4cm}&&u(\xi)+\frac{\beta (t)}{\alpha (t)\dot{\beta}(t)}\left[%
\beta (t)u(\xi)-\dot{\alpha}(t)\xi\right]\frac{du(\xi)}{d\xi}=  \notag \\
\hspace{-0.4cm}&&-2u_0\frac{\gamma (t)}{\alpha (t)\dot{\beta}(t)}\frac{%
d\theta (\xi)}{d\xi}-\frac{G\eta (t)}{\alpha ^2(t)\dot{\beta}(t)}\frac{m(\xi)%
}{\xi ^2}+\frac{2}{3}\Omega ^2(t)\frac{\alpha (t)}{\dot{\beta}(t)}\xi.
\notag \\
\end{eqnarray}

The existence of a similarity solution requires that the condition
\begin{equation}
\dot{\alpha}(t)=\beta(t),
\end{equation}
must be satisfied by the functions $\alpha (t)$ and $\beta (t)$. The
explicit time dependence in the evolution equations of the condensate dark
matter halo could be removed if the conditions
\begin{equation}  \label{68}
\frac{\gamma (t)\alpha ^3(t)}{\eta (t)}=c_1,\frac{\gamma (t)}{\dot{\eta}(t)}%
\alpha ^2(t)\dot{\alpha}(t)=c_2, \frac{\dot{\alpha}^2(t)}{\alpha (t)\ddot{%
\alpha }(t)}=c_3,
\end{equation}
\begin{equation}  \label{69a}
2u_0\frac{\gamma (t)}{\alpha (t)\ddot{\alpha}(t)}=c_4,\frac{G\eta (t)}{%
\alpha ^2(t)\ddot{\alpha}(t)}=c_5,\frac{2}{3}\Omega ^2(t)\frac{\alpha (t)}{%
\ddot{\alpha}(t)}=c_6,
\end{equation}
are also satisfied, where $c_i$, $i=1,...,6$ are constants. The third
relation in Eqs.~(\ref{68}) determines the function $\alpha (t)$ as a
solution of the second order differential equation
\begin{equation}
\ddot{\alpha}(t)=\frac{\dot{\alpha }^2(t)}{c_3\alpha (t)},
\end{equation}
with a particular solution given by
\begin{equation}  \label{A9}
\alpha (t)=C_1t^n,
\end{equation}
where $C_1$ is an arbitrary constant of integration, and $%
n=c_3/\left(c_3-1\right)$. The first two of the equations (\ref{69a}) give
\begin{equation}
\gamma (t)=\frac{c_4}{2u_0}\alpha (t)\ddot{\alpha}(t), \eta (t)=\frac{c_5}{G}%
\alpha ^2(t)\ddot{\alpha }(t).
\end{equation}
The substitution of the above relations into the first of Eqs.~(\ref{68})
gives $\alpha ^2(t)=\mathrm{constant}$, a result that obviously contradicts
Eq.~(\ref{A9}). Hence the explicit time dependence in the hydrodynamic
equations describing the dynamics of a polytropic fluid with equation of
state $p \sim \rho ^{\gamma}$, $\gamma =2$, cannot be removed by means of
the similarity transformations (\ref{A1}), and therefore there are no
self-similar solutions describing the evolution of Bose-Einstein Condensate
dark matter halos. However, for polytropic fluids with $\gamma \neq 2$ the
hydrodynamic type evolution equations always admit self-similar solutions.


\begin{thebibliography}{999}

\bibitem{d1} L. E. Strigari, Physics Reports \textbf{531}, 1 (2013).

\bibitem{d2} K. M. Zurek, Physics Reports \textbf{537}, 91 (2014).

\bibitem{d3} S. Courteau et al., Reviews of Modern Physics \textbf{86}, 47
(2014).

\bibitem{d4} H. Baer, K.-Y. Choi, J. E. Kim, and L. Roszkowski, Physics
Reports \textbf{555}, 1 (2015).

\bibitem{d5} T. Aramaki et al., Physics Reports \textbf{618}, 1 (2016).

\bibitem{d6} M. R. Buckley and  A. H. G. Peter, Physics Reports {\bf  761}, 1 (2018).

\bibitem{d7} S. Tulin and H.-B. Yu, Physics Reports {\bf 730},  1 (2018).

\bibitem{Rubin} V. Rubin, W. K.  Thonnard, Jr., and  N. Ford, The Astrophysical Journal {\bf  238},  471 (1980).

\bibitem{Persic} M. Persic, P. Salucci, and F. Stel, Monthly Notices of the
Royal Astronomical Society \textbf{458}, 4172 (1996).

\bibitem{Read} J. I. Read, G. Iorio, O. Agertz, and F. Fraternali, Monthly
Notices of the Royal Astronomical Society \textbf{462}, 3628 (2016).

\bibitem{Salucci} E. V. Karukes and P. Salucci, arXiv:1609.06903
[astro-ph.GA] (2016).

\bibitem{Haghi} H. Haghi, A. E. Bazkiaei, A. Hasani Zonoozi, and P. Kroupa,
Monthly Notices of the Royal Astronomical Society \textbf{458}, 4172 (2016).

\bibitem{BT1} J. Binney and S. Tremaine S, Galactic dynamics, Princeton University Press, Princeton, N. J.; Woodstock, 2008

\bibitem{20}  A. V. Kravtsov and S. Borgani, Annual Review of Astronomy and Astrophysics {\bf 50}, 353 (2012).

\bibitem{Planck1} P. A. R. Ade et al., Astronomy and Astrophysics \textbf{594}%
, A13 (2016).

\bibitem{Wegg} C. Wegg, O. Gerhard, and M. Portail, Monthly Notices of the
Royal Astronomical Society \textbf{463}, 557 (2016).

\bibitem{Munoz} J. B. Mu$\tilde{\mathrm{n}}$oz, E. D. Kovetz, L. Dai, and M.
Kamionkowski, Phys. Rev. Lett. \textbf{117}, 091301 (2016).

\bibitem{Chuda} A. Chudaykin, D. Gorbunov, and I. Tkachev, Phys. Rev.
\textbf{D 94}, 023528 (2016).

\bibitem{massey2007dark} R. Massey et al., Nature \textbf{445}, 286 (2007).

\bibitem{Overduin} J. M. Overduin and P. S. Wesson, Physics Reports \textbf{%
402}, 267 (2004).

\bibitem{Cui} Y. Cui, Modern Physics Letters \textbf{A 30}, 1530028 (2015).

\bibitem{Matsumoto} S. Matsumoto, S. Mukhopadhyay, and Y.-L. Sming Tsai,
Phys. Rev. \textbf{D 94}, 065034 (2016).

\bibitem{Mielke} D. Casta$\tilde{\mathrm{n}}$eda Valle and E. W. Mielke,
Physics Letters \textbf{B 758}, 93 (2016).

\bibitem{Schwabe} B. Schwabe, J. C. Niemeyer, and J. F. Engels, Phys. Rev.
\textbf{D 94}, 043513 (2016).

\bibitem{Milgrom} M. Milgrom, Astrophys. J. \textbf{270}, 365 (1983).

\bibitem{alt1} M. K. Mak and T. Harko, Phys. Rev. \textbf{D 70}, 024010
(2004).

\bibitem{alt2} T. Harko and K. S. Cheng, Phys. Rev. \textbf{D 76}, 044013
(2007).

\bibitem{alt3} O. Bertolami, C. G. Boehmer, T. Harko, and F. S. N. Lobo,
Phys. Rev. \textbf{D 75}, 104016 (2007).

\bibitem{alt4} C. G. Boehmer, T. Harko, and F. S. N. Lobo, JCAP \textbf{0803}%
, 024 (2008).

\bibitem{alt5} H. R. Sepangi and S. Shahidi, Class. Quant. Grav. \textbf{38}
26, 185010 (2009).

\bibitem{alt6} A. S. Sefiedgar, K. Atazadeh, and H. R. Sepangi, Phys. Rev.
\textbf{D 80}, 064010 (2009).

\bibitem{alt7} A. S. Sefiedgar, Z. Haghani, and H. R. Sepangi, Phys. Rev.
\textbf{D 85}, 064012 (2012).

\bibitem{alt8} O. Bertolami, P. Frazao, and J. Paramos, Phys. Rev. \textbf{D
86}, 044034 (2012).

\bibitem{alt9} L. Lombriser, F. Schmidt, T. Baldauf, R. Mandelbaum, U.
Seljak, and R. E. Smith, Phys. Rev. \textbf{D 85}, 102001 (2012).

\bibitem{alt10} S. Capozziello, T. Harko, T. S. Koivisto, F. S. N. Lobo, and
G. J. Olmo, JCAP \textbf{07} 024 (2013).

\bibitem{alt11} T. Harko, F. S. N. Lobo, M. K. Mak, and S. V. Sushkov, Mod.
Phys. Lett. \textbf{A 29}, 1450049 (2014).

\bibitem{alt12} R. Myrzakulov, L. Sebastiani, S. Vagnozzi, and S. Zerbini, Class. Quant. Grav. {\bf 33}, 125005 (2016).

\bibitem{alt13} L. Sebastiani, S. Vagnozzi, and R. Myrzakulov, Adv. High Energy Phys. {\bf 2017},  3156915 (2017).

\bibitem{alt14} S. Vagnozzi, Class. Quant. Grav. {\bf 34},  185006 (2017).

\bibitem{book} T. Harko and F. S. N. Lobo, Extensions of f(R) Gravity:
Curvature-Matter Couplings and Hybrid Metric-Palatini Theory, Cambridge University Press, Cambridge, 2018

\bibitem{Chann1} M. Aguilar et al., Phys. Rev. Lett. \textbf{113}, 221102
(2014).

\bibitem{Chann2} F. Calore et al., JCAP \textbf{03}, 38 (2015).

\bibitem{Chann3} T. Daylan et al., Physics of the Dark Universe \textbf{12},
1 (2016).

\bibitem{Chan1} M. H. Chan, Phys. Rev. \textbf{D 94}, 023507 (2016).

\bibitem{Chan2} M. H. Chan, Astrophys. J. \textbf{844}, 9 (2017).

\bibitem{Chan3} M. H. Chan, Phys. Rev. \textbf{D 96}, 043009 (2017).

\bibitem{Chan4} M. H. Chan and C. H. Leung, Nature Scientific Reports
\textbf{7}, 14895 (2017).

\bibitem{Chan5} M. H. Chan, Monthly Notices of the Royal Astronomical
Society \textbf{474}, 2576 (2018).

\bibitem{NFW} J. F. Navarro, C. S. Frenk, and S. D. M. White, Astrophys. J.
\textbf{462}, 563 (1996).

\bibitem{H} A. Genina, A. Benitez-Llambay, C. S. Frenk, S. Cole, A. Fattahi,
J. F. Navarro, K. A. Oman, T. Sawala, and T. Theuns, Monthly Notices of the
Royal Astronomical Society \textbf{474}, 1398 (2018).

\bibitem{OH} M. Arca-Sedda and R. Capuzzo-Dolcetta, Monthly Notices of the
Royal Astronomical Society \textbf{464}, 3060 (2017).

\bibitem{Boylan1} M. Boylan-Kolchin, J. S. Bullock, and M. Kaplinghat,
Monthly Notices of the Royal Astronomical Society \textbf{415}, L40 (2011).

\bibitem{Boylan2} M. Boylan-Kolchin, J. S. Bullock, and M. Kaplinghat,
Monthly Notices of the Royal Astronomical Society \textbf{422}, 1203 (2012).

\bibitem{si1} E. D. Carlson, M. E.  Machacek, and L. J.  Hall,  1992, Astrophys. J.  {\bf 398}, 43 (1992).

\bibitem{si2} D. N. Spergel and P. J. Steinhardt, Phys. Rev. Lett. {\bf 84}, 3760 (2000).

\bibitem{si3} C. Firmani, E. D’Onghia, V. Avila-Reese, G. Chincarini, and X. Hernández,  MNRAS {\bf 315}, L29 (2000).

\bibitem{si4} R. Dav\'{e}, D. N. Spergel, P. J.  Steinhardt, and B. D. Wandelt,  Astrophys. J.  {\bf 547}, 574 (2001).

\bibitem{si5} A. Kamada, M. Kaplinghat, A. B.  Pace, and H.-B. Yu,   Phys. Rev. Lett. {\bf 119}, 111102 (2017).

\bibitem{si6} P. Creasey, O. Sameie, L. V. Sales,  et al., MNRAS {\bf 468}, 2283 (2017).

\bibitem{si7} T. Ren, A. Kwa, M. Kaplinghat, and H. B. Yu,  arXiv: 1808.05695 (2018).

\bibitem{si8} A. Robertson, R. Massey, V. Eke,  et al., MNRAS {\bf 476}, L20  (2018).

\bibitem{Harvey} D. Harvey, R. Massey, T. Kitching, A. Taylor, and E.
Tittley, Science \textbf{347}, 1462 (2015).

\bibitem{Jauzac} M. Jauzac et al., Monthly Notices of the Royal Astronomical
Society \textbf{463}, 3876 (2016).

\bibitem{Carlson} E. D. Carlson, M. E. Machacek, and J. L. Hall, Astrophys.
J. \textbf{398}, 43 (1992).

\bibitem{Laix} A. A. de Laix, R. J. Scherrer, and R. K. Schaefer, Astrophys.
J. \textbf{452}, 495 (1995).

\bibitem{Saxton1} C. J. Saxton and I. Ferreras, Monthly Notices of the Royal
Astronomical Society \textbf{405}, 77 (2010).

\bibitem{Saxton2} C. J. Saxton, Monthly Notices of the Royal Astronomical
Society \textbf{430}, 1578 (2013).

\bibitem{Dooley} G. A. Dooley, A. H. G. Peter, M. Vogelsberger, J. Zavala,
and A. Frebel, Monthly Notices of the Royal Astronomical Society \textbf{461}%
, 710 (2016).

\bibitem{Elbert} O. D. Elbert, J. S. Bullock, M. Kaplinghat, S.
Garrison-Kimmel, A. S. Graus, and M. Rocha, arXiv:1609.08626 [astro-ph.GA]
(2016).

\bibitem{Bose} S. N. Bose, Z. Phys. \textbf{26}, 178 (1924).

\bibitem{Ein} A. Einstein, Sitzungsberichte der Preussischen Akademie der
Wissenschaften, Physikalisch-mathematische Klasse, \textbf{1924}, 261 (1924).

\bibitem{Ein1} A. Einstein, Sitzungsberichte der Preussischen Akademie der
Wissenschaften, Physikalisch-mathematische Klasse, \textbf{1925}, 3 (1925).

\bibitem{Dalfovo} F. Dalfovo, S. Giorgini, L. P. Pitaevskii, and S.
Stringari, Rev. Mod. Phys. \textbf{71}, 463 (1999).

\bibitem{Pita} L. Pitaevskii and S. Stringari, Bose-Einstein condensation,
Clarendon Press, Oxford (2003).

\bibitem{Pethick} C. J. Pethick and H. Smith, Bose-Einstein condensation in
dilute gases, Cambridge, Cambridge University Press (2008).

\bibitem{ZNG} A. Griffin, T. Nikuni, and E. Zaremba, Bose-condensed gases at
finite temperatures, Cambridge, Cambridge University Press, (2009).

\bibitem{exp1} M. H. Anderson, J. R. Ensher, M. R. Matthews, C. E. Wieman,
and E. A. Cornell, Science \textbf{269}, 198 (1995).

\bibitem{exp2} C. C. Bradley, C. A. Sackett, J. J. Tollett, and R. G. Hulet,
Phys. Rev. Lett. \textbf{75}, 1687 (1995).

\bibitem{exp3} K. B. Davis, M. O. Mewes, M. R. Andrews, N. J. van Drutten,
D. S. Durfee, D. M. Kurn, and W. Ketterle, Phys. Rev. Lett. \textbf{75},
3969 (1995).

\bibitem{starsm1} M. Membrado, J. Abad, A. F. Pacheco and J. Sanudo, Phys.
Rev. \textbf{D 40}, 2736 (1989).

\bibitem{stars0} X. Z. Wang, Phys. Rev. \textbf{D 64}, 124009 (2001).

\bibitem{stars1} P. H. Chavanis and T. Harko, Phys. Rev. \textbf{D 86},
064011 (2012).

\bibitem{stars2} X. Y. Li, T. Harko, and K. S. Cheng, JCAP \textbf{06},001
(2012).

\bibitem{stars3} X. Y. Li, F. Wang, and K. S. Cheng, JCAP \textbf{10}, 031
(2012).

\bibitem{stars4} P. H. Chavanis, Eur. Phys. J. Plus \textbf{130}, 181 (2015).

\bibitem{stars5} S. Latifah, A. Sulaksono, and T. Mart, Phys. Rev. \textbf{D
90}, 127501 (2014).

\bibitem{stars6} A. Mukherjee, S. Shah, and S. Bose, Phys. Rev. \textbf{D 91}%
, 084051 (2015).

\bibitem{stars7} B. Danila, T. Harko, and Z. Kovacs, Eur. Phys. J. \textbf{C
75}, 203 (2015).

\bibitem{stars8} N. Kan and K. Shiraishi, Phys. Rev. \textbf{D 94}, 104042
(2016).

\bibitem{stars9} D. Croon, J. J. Fan, and C. Sun, Journal of Cosmology and Astroparticle Physics {\bf 04}, 008 (2019).

\bibitem{early1} M. Membrado, A. F. Pacheco, and J. S. Sanudo, Astron.
Astrophys. \textbf{217}, 92 (1989).

\bibitem{early2} J. Sin, Phys. Rev. \textbf{D 50}, 3650 (1994).

\bibitem{early3} S. U. Ji and S. J. Sin, Phys. Rev. \textbf{D50}, 3655
(1994).

\bibitem{early4} M. Membrado and J. A. L. Aguerri, Int. J. Mod. Phys.
\textbf{D 5}, 257 (1996).

\bibitem{early5} M. Membrado, Monthly Notices of the Royal Astronomical
Society \textbf{296}, 21 (1998).

\bibitem{early6} W. Hu, R. Barkana, and A. Gruzinov, Phys. Rev. Lett.
\textbf{85} 1158, (2000).

\bibitem{early7} J. Goodman, New Astronomy \textbf{5}, 103 (2000)

\bibitem{early8} P. J. E. Peebles, Astrophys. J. \textbf{534}, L127 (2000).

\bibitem{early9a} K. R. W. Jones and D. Bernstein, Classical and Quantum
Gravity \textbf{18}, 1513 (2001).

\bibitem{rotha2002vortices} R. P. Yu and M. J. Morgan, Classical and Quantum
Gravity \textbf{19}, L157, (2002).

\bibitem{silverman2002dark} M. P. Silverman and R. L. Mallett, General
Relativity and Gravitation \textbf{34}, 633 (2002).

\bibitem{early9} A. Arbey, J. Lesgourgues and P. Salati, Phys. Rev. \textbf{%
D 68}, 023511 (2003).

\bibitem{GP1} E. P. Gross, Phys. Rev. \textbf{106}, 161 (1957).

\bibitem{GP2} V. L. Ginzburg and L. P. Pitaevskii, Sov. Phys. JETP \textbf{7}%
, 858 (1958).

\bibitem{BoHa07a} C. G. Boehmer and T. Harko, JCAP \textbf{06}, 025 (2007).

\bibitem{Horedt} G. P. Horedt, Polytropes: Applications in Astrophysics and
Related Fields, Kluwer Academic Publishes, New York, Boston, Dordrecht, 2004

\bibitem{HaM} T. Harko and E. J. M. Madarassy, JCAP \textbf{01}, 020 (2011).

\bibitem{Har1} T. Harko, JCAP \textbf{1105}, 022 (2011).

\bibitem{inv0} P. Sikivie and Q. Yang, Phys. Rev. Lett. \textbf{103}, 111301
(2009).

\bibitem{inv1} J.-W.Lee, Phys. Lett. \textbf{B 681}, 118 (2009).

\bibitem{inv2} J.-W. Lee and S. Lim, JCAP \textbf{1001}, 007 (2010).

\bibitem{inv3} B. Kain and H. Y. Ling, Phys. Rev. \textbf{D 82}, 064042
(2010).

\bibitem{inv4} N. T. Zinner, Physics Research International \textbf{2011},
734543 (2011).

\bibitem{inv5} P.-H. Chavanis, Phys. Rev. \textbf{D 84}, 043531 (2011).

\bibitem{inv6} P.-H. Chavanis and L. Delfini, Phys. Rev. \textbf{D 84},
043532 (2011).

\bibitem{inv7} P.-H. Chavanis, Phys. Rev. \textbf{D 84}, 063518 (2011).

\bibitem{inv8} P.-H. Chavanis, Phys. Rev. \textbf{E 84}, 031101 (2011).

\bibitem{inv9} T. Harko, Monthly Notices of the Royal Astronomical Society
\textbf{413}, 3095 (2011).

\bibitem{inv10} V. H. Robles and T. Matos, Mon. Monthly Notices of the Royal
Astronomical Society \textbf{422}, 282 (2012).

\bibitem{inv11} M. O. C. Pires and J. C. C. de Souza, JCAP \textbf{11}, 024
(2012).

\bibitem{inv12} P.-H. Chavanis, Astron. Astrophys. \textbf{537}, A127 (2012).

\bibitem{inv13} H. Velten and E. Wamba, Phys. Lett. \textbf{B 709}, 1 (2012).

\bibitem{inv14} T. Harko and G. Mocanu, Phys. Rev. \textbf{D 85}, 084012
(2012).

\bibitem{inv15} T. Rindler-Daller and P. R. Shapiro, Monthly Notices of the
Royal Astronomical Society \textbf{422}, 135 (2012).

\bibitem{inv16} F. S. Guzman, F. D. Lora-Clavijo, J. J. Gonzalez-Aviles, and
F. J. Rivera-Paleo, Phys. Rev. \textbf{D 89}, 063507 (2014).

\bibitem{inv17} T. Rindler-Daller and P. R. Shapiro, Mod. Phys. Lett.
\textbf{A 29}, 1430002 (2014).

\bibitem{inv18} R. C. Freitas and S. V. B. Goncalves, JCAP \textbf{04}, 049
(2013).

\bibitem{inv19} E. J. M. Madarassy and V. T. Toth, Computer Physics
Communications \textbf{184}, 1339 (2013).

\bibitem{inv20} F. S. Guzman, F. D. Lora-Clavijo, J. J. Gonzalez-Aviles, and
F. J. Rivera-Paleo, JCAP \textbf{09}, 034 (2013).

\bibitem{inv21} J. C. C. de Souza and M. O. C. Pires, JCAP \textbf{03}, 010
(2014).

\bibitem{inv22} V. T. Toth, arXiv:1402.0600 (2014).

\bibitem{inv23} T. Harko, Phys. Rev. \textbf{D 89}, 084040 (2014).

\bibitem{inv24} M. Dwornik, Z. Keresztes, and L. A. Gergely, Chapter 6 of
``Recent Development in Dark Matter Research", Eds. N. Kinjo, A. Nakajima,
Nova Science Publishers (2014), arXiv:1312.3715.

\bibitem{inv25} M.-H. Li and Z.-B. Li, Phys. Rev. \textbf{D 89}, 103512
(2014).

\bibitem{inv26} T. Harko and M. J. Lake, Phys. Rev. \textbf{D 91}, 045012
(2015).

\bibitem{inv27} E. J. M. Madarassy and V. T. Toth, Phys. Rev. \textbf{D 91},
044041 (2015).

\bibitem{inv28} J. C. C. de Souza and M. Ujevic, Gen. Rel. Grav. \textbf{47}%
, 100 (2015).

\bibitem{inv29} T. Harko and F. S. N. Lobo, Phys. Rev. \textbf{D 92}, 043011
(2015).

\bibitem{inv30} V. H. Robles, V. Lora, T. Matos, and F. J. Sanchez-Salcedo,
Astrophys. J. \textbf{810}, 99 (2015).

\bibitem{inv31} T. Harko, P.-X Liang, S.-D. Liang, and G. Mocanu, JCAP
\textbf{11}, 027 (2015).

\bibitem{inv32} L. A. Martinez-Medina, H. L. Bray and T. Matos, JCAP \textbf{%
12}, 025 (2015).

\bibitem{inv33} K. Schroven, M. List, and C. L\"{a}mmerzahl, Phys. Rev. {\bf D 92}, 124008 (2015).

\bibitem{inv34} H.-Y. Schive, T. Chiueh, T. Broadhurst, and K.-W. Huang, The
Astrophysical Journal \textbf{818}, 89 (2016).

\bibitem{inv35} S.-R. Chen, H.-Y. Schive, T. Chiueh, Monthly Notices of the
Royal Astronomical Society \textbf{468}, 1338 (2017).

\bibitem{inv36} E. Calabrese and D. N. Spergel, Monthly Notices of the Royal
Astronomical Society \textbf{460}, 4397 (2016).

\bibitem{inv37} P.-H. Chavanis, Phys. Rev. {\bf D 94}, 083007 (2016).

\bibitem{inv38} P. S. Bhupal Dev, M. Lindner, and S. Ohmer,Phys. Lett.
\textbf{B 773}, 219 (2017).

\bibitem{inv39} S. Sarkar, C. Vaz, and L. C. R. Wijewardhana, Phys. Rev. {\bf D 97}, 103022 (2018).

\bibitem{inv40} W.-J. Chung and L. Nelson, Journal of Cosmology and Astroparticle Physics {\bf 06}, 010 (2018).

\bibitem{inv41} X. Zhang, M. H. Chan, T. Harko, S.-D. Liang, and C. S. Leung, The European Physical Journal {\bf C 78}, 346 (2018).

\bibitem{inv41a} E. Kun, Z. Keresztes, S. Das, and L. \'{A}. Gergely, Symmetry {\bf 10}, 520 (2018).

\bibitem{inv42} V. Sreenath, Phys. Rev. {\bf D 99}, 043540 (2019).

\bibitem{inv43} J.-W. Lee, H.-C. Kim, and J. Lee, 	arXiv:1901.00305 (2019).

\bibitem{inv44} S. HajiSadeghi, S. Smolenski, and J. Wudka, Phys. Rev. {\bf D 99}, 023514 (2019).

\bibitem{inv45} M. Morikawa and S. Takahashi, arXiv:1903.02986 (2019).

\bibitem{Hui} L. Hui, J. P. Ostriker, S. Tremaine, and E. Witten, Phys. Rev.
\textbf{D 95}, 043541 (2017).

\bibitem{Jeans} J. H. Jeans, Philos. Trans. R. Soc. London, Ser. A {\bf 199}, 1 (1902).

\bibitem{Chand1} S. Chandrasekhar, Proceedings of the Royal Society of London. Series A, Mathematical and Physical Sciences {\bf 210},  26 (1951).

\bibitem{Chand2} S. Chandrasekhar, Astrophysical Journal {\bf 119}, 7 (1954).

\bibitem{Chand2a} W. Fricke, Astrophysical Journal {\bf 120}, 356 (1954).

\bibitem{Chand3} S. Chandrasekhar, Vistas in Astronomy {\bf 1}, 344 (1955).

\bibitem{Chand4} N. Bel and E. Schatzman, Reviews of Modern Physics {\bf 30}, 1015 (1958).

\bibitem{Chand5} I. L. Genkin, Soviet Astronomy {\bf 12}, 1004 (1969).

\bibitem{Chand6} A. P. Boss, Astrophysical Journal {\bf 244}, 40 (1981).

\bibitem{Chand7} V. M. Cadez, Astronomy and Astrophysics {\bf 235}, 242 (1990).

\bibitem{Chand8} B. G. Elmegreen, The Astrophysical Journal {\bf 433}, 39 (1994).

\bibitem{Chand9} T. A. Thompson, The Astrophysical Journal {\bf 684}, 212 (2008).

\bibitem{Chand10} G.-X. Li, Monthly Notices of the Royal Astronomical Society {\bf 465}, 667 (2017).

\bibitem{kin1} D. Lynden-Bell, Mon. Not. R. Astron. Soc. {\bf 124}, 279 (1962).

\bibitem{kin2} C. Simon, Bull. Acad. R. Belg. Soc. {\bf 97}, 7 (1962).

\bibitem{kin3} P. A. Sweet, Mon. Not. R. Astron. Soc. {\bf 125}, 285 (1963).

\bibitem{kin4} S. A. Trigger, A. I. Ershkovich, G. J. F. van Heist, P. P. J. M. Schram, Phys. Rev.  {\bf E 69}, 066403 (2004).

\bibitem{kin5} P. H. Chavanis, Eur. Phys. J.  {\bf B 85}, 229 (2012).

\bibitem{kin6} A. Su\'{a}rez and P.-H. Chavanis, Phys. Rev. {\bf D 98}, 083529 (2018).

\bibitem{CollBEC1}  E. A. Donley, N. R. Claussen, S. L. Cornish, J. L.
Roberts, E. A. Cornell, and C. E. Wieman, Nature {\bf 412},
295 (2001).

\bibitem{CollBEC2}  J. Sopik, C. Sire, and P.-H. Chavanis, Phys. Rev. {\bf E 74},
011112 (2006).

\bibitem{CollBEC3}  P.-H. Chavanis, Phys. Rev. {\bf E 84}, 031101 (2011).

\bibitem{CollBEC4} P.-H. Chavanis and C. Sire, Phys. Rev. {\bf E 83}, 031131 (2011).

\bibitem{Lush} P. M. Lushnikov, Phys. Rev. {\bf A 82}, 023615 (2010).

\bibitem{Bos}  T. Fukuyama, M. Morikawa and T. Tatekawa, Journal of Cosmology and Astroparticle Physics {\bf 06}, 033 (2008).

\bibitem{var} V. M. Perez-Garcia, H. Michinel, J. I. Cirac, M. Lewenstein,
and P. Zoller, Phys. Rev.  \textbf{A 56}, 1424 (1997).

\bibitem{vor0} A. Fetter, Journal of Low Temperature Physics {\bf 161}, 445 (2010).

\bibitem{vor1} M. Caracanhas, A. L. Fetter, S. R. Muniz, K. M. F. Magalhaes, G. Roati, G. Bagnato, and V. S. Bagnato, Journal of Low Temperature Physics {\bf 166}, 49 (2012).

\bibitem{vor2} M. Caracanhas, A. L.  Fetter, G. Baym, S. R.  Muniz, and V. S.  Bagnato, Journal of Low Temperature Physics {\bf 170}, 133 (2013).

\bibitem{Chandb} S. Chandrasekhar, An Introduction to the Study of Stellar
Structure, Univ. of Chicago Press, Chicago, U. S. A., 1939

\bibitem{Shap} S. L. Shapiro and S. A. Teukolsky, Black Holes, White Dwarfs
and Neutron Stars: The Physics of Compact Objects, John Wiley \& Sons, Inc.,
New York, U. S. A., 1983

\bibitem{Kipp} R. Kippenhahn, A. Weigert, and A. Weiss, Stellar Structure and Evolution, Springer-Verlag, Berlin, Heidelberg, 2012

\bibitem{turb1} R. J. Donnelly and C. E.  Swanson,  Journal of Fluid Mechanics {\bf 173}, 387429 (1986).

\bibitem{turb2} L. Madeira, M. A. Caracanhas, F. E. A. dos Santos, and V. S. Bagnato, arXiv:1903.12215 [cond-mat.quant-gas] (2019).

\bibitem{SP1} L. M. Widrow and N. Kaiser,  Astrophys. J. {\bf 416},  L71 (1993).

\bibitem{SP2} H. Y. Schive, T. Chiueh and T. Broadhurst,  Nature Phys. {\bf 10},  496 (2014).

\bibitem{SP3} H. Y. Schive, M. H. Liao, T. P. Woo, S. K. Wong, T. Chiueh, T. Broadhurst and W.-
Y. P. Hwang,  Phys. Rev. Lett. {\bf 113},  261302 (2014).

\bibitem{SP4} B. Schwabe, J. C. Niemeyer and J. F. Engels,  Phys. Rev. {\bf D 94},  043513 (2016).

\bibitem{SP5} M. Garny and T. Konstandin, Journal of Cosmology and Astroparticle Physics {\bf 01}, 009 (2018).

\bibitem{hom1} P. Goldreich and S. V. Weber, Astrophys. J. \textbf{238}, 991
(1980).

\bibitem{hom2} A. Yahil, Astrophysical Journal {\bf 265},  1047 (1983).

\bibitem{hom3} P. Blottiau, J. P.  Chieze, and S. Bouquet, Astronomy and Astrophysics {\bf  207}, 24 (1988).

\bibitem{hom4} Y.-Q. Lou and C.-H. Shi, Monthly Notices of the Royal Astronomical Society {\bf 445}, 1186 (2014).

\bibitem{hom5} D. L. Li, Y.-Q. Lou, and J. Esimbek, Monthly Notices of the Royal Astronomical Society {\bf 473}, 2441 (2018).

\end{thebibliography}
\end{document}